\documentclass[reqno,a4paper,11pt]{article}
\pdfoutput=1
\usepackage{xcolor}

\usepackage{graphicx}
\usepackage[textwidth = 430 pt, textheight = 630 pt]{geometry}

\definecolor{MyDarkBlue}{rgb}{0.15,0.25,0.45}
\usepackage{epsfig,rotating}
\usepackage{amsmath,amssymb}
\usepackage{amsfonts}
\usepackage{mathrsfs}
\usepackage{bbm}
\usepackage[normalem]{ulem}
\usepackage[T1]{fontenc}

\usepackage{latexsym}
\usepackage{amsthm}
\usepackage[all,knot]{xy}
\xyoption{arc}

\usepackage[utf8x]{inputenc}

\usepackage{hyperref}
\hypersetup{
hypertexnames=false,
colorlinks=true,
citecolor=MyDarkBlue,
linkcolor=MyDarkBlue,
urlcolor=MyDarkBlue,
pdfauthor={Andreas Deser, Marc Andre Heller and Christian Saemann},
pdftitle={Extended Riemannian Geometry II: Local Heterotic Double Field Theory},
pdfsubject={hep-th math-ph}
breaklinks=true
}

\usepackage{tikz}
\usepackage{mathtools}

\let\fn\footnote
\renewcommand{\footnote}[1]{\linespread{1.1}\fn{#1}\linespread{1.29}}

\makeatletter\renewcommand{\section}{\@startsection
{section}{1}{\z@}{-3.5ex plus -1ex minus
    -.2ex}{2.3ex plus .2ex}{\bf }}
\makeatletter\renewcommand{\subsection}{\@startsection{subsection}{2}{\z@}{-3.25ex
plus -1ex minus
   -.2ex}{1.5ex plus .2ex}{\bf }}
\makeatletter\renewcommand{\subsubsection}{\@startsection{subsubsection}{3}{-2.45ex}{-3.25ex
plus -1ex minus -.2ex}{1.5ex plus .2ex}{\it }}
\renewcommand{\thesection}{\arabic{section}}
\renewcommand{\thesubsection}{\arabic{section}.\arabic{subsection}}
\renewcommand{\@seccntformat}[1]{\@nameuse{the#1}.~~}

\renewcommand{\theequation}{\thesection.\arabic{equation}}
\makeatletter \@addtoreset{equation}{section}

\usepackage[toc,page]{appendix}

\newtheorem{thm}{Theorem}[section]
\renewcommand{\thethm}{\thesection.\arabic{thm}}

\newtheorem{remark}[thm]{Remark}

\renewcommand{\appendices}{
\section*{Appendix}\label{appendices}\setcounter{subsection}{0}
\addcontentsline{toc}{section}{Appendix}
\setcounter{equation}{0}
\makeatletter
\renewcommand{\theequation}{\Alph{subsection}.\arabic{equation}}
\renewcommand{\thesubsection}{\Alph{subsection}}
\renewcommand{\thethm}{\Alph{subsection}.\arabic{thm}}
\@addtoreset{equation}{subsection}
\@addtoreset{thm}{subsection}
\makeatother
}

\def\slasha#1{\setbox0=\hbox{$#1$}#1\hskip-\wd0\hbox to\wd0{\hss\sl/\/\hss}}

\def\periodb#1{\setbox0=\hbox{$#1$}#1\hskip-\wd0\hbox to\wd0{-}}

\newcommand{\unit}{\mathbbm{1}}   			

\newcommand{\CB}{\mathcal{B}}

\newcommand{\CC}{\mathcal{C}}
\newcommand{\CCC}{\mathscr{C}}

\newcommand{\CF}{\mathcal{F}}

\newcommand{\CCG}{\mathscr{G}}
\newcommand{\CH}{\mathcal{H}}

\newcommand{\CL}{\mathcal{L}}
\newcommand{\CM}{\mathcal{M}}

\newcommand{\CQ}{\mathcal{Q}}

\newcommand{\CR}{\mathcal{R}}

\newcommand{\CT}{\mathcal{T}}

\newcommand{\CV}{\mathcal{V}}

\newcommand{\CW}{\mathcal{W}}

\newcommand{\CCX}{\mathscr{X}}

\newcommand{\CE}{\mathcal{E}}
\newcommand{\frg}{\mathfrak{g}}				
\newcommand{\frh}{\mathfrak{h}}				

\newcommand{\FR}{\mathbbm{R}}     			
\newcommand{\NN}{\mathbbm{N}}     			

\newcommand{\dd}{\mathrm{d}}     			
\newcommand{\dpar}{\partial}     			

\newcommand{\de}{\mathrm{e}}     			
\newcommand{\eps}{{\varepsilon}}			

\newcommand{\eand}{{\qquad\mbox{and}\qquad}}     		
\newcommand{\ewith}{{\qquad\mbox{with}\qquad}}

\newcommand{\der}[1]{\frac{\dpar}{\dpar #1}}   		
\newcommand{\tr}{\,\mathrm{tr}\,}     			
\newcommand{\ad}{\mathrm{ad}}     			

\newcommand{\fd}{\mathfrak{d}}

\newcommand{\au}{\mathfrak{u}}

\newcommand{\astring}{\mathfrak{string}}
\newcommand{\aso}{\mathfrak{so}}
\newcommand{\ao}{\mathfrak{o}}
\newcommand{\apin}{\mathfrak{pin}}
\renewcommand{\ae}{\mathfrak{e}}

\newcommand{\aspin}{\mathfrak{spin}}

\newcommand{\sU}{\mathsf{U}}     			

\newcommand{\sG}{\mathsf{G}}
\newcommand{\sL}{\mathsf{L}}
\newcommand{\sT}{\mathsf{T}}

\newcommand{\sLie}{\mathsf{Lie}}

\newcommand{\sO}{\mathsf{O}}

\newcommand{\sSO}{\mathsf{SO}}
\newcommand{\sSpin}{\mathsf{Spin}}
\newcommand{\sString}{\mathsf{String}}
\newcommand{\sE}{\mathsf{E}}

\newcommand{\acton}{\vartriangleright}     			
\renewcommand{\remark}[1]{}     				
\def\tyng(#1){\hbox{\tiny$\yng(#1)$}}			
\def\tyoung(#1){\hbox{\tiny$\young(#1)$}}			

\newcommand{\beq}{\begin{eqnarray}}
\newcommand{\eeq}{\end{eqnarray}}

\begin{document}
\begin{titlepage}
\begin{flushright}
 EMPG--17--18
\end{flushright}
\vskip 1.0cm
\begin{center}
{\LARGE \bf Extended Riemannian Geometry II:\\[0.2cm] Local Heterotic Double Field Theory}
\vskip 1.cm
{\Large Andreas Deser$^{a}$, Marc Andre Heller$^{b}$ and Christian S\"amann$^c$}
\setcounter{footnote}{0}
\renewcommand{\thefootnote}{\arabic{thefootnote}}
\vskip 1cm
{\em${}^a$ Istituto Nationale di Fisica Nucleare \\
Via P.~Giuria 1\\
10125 Torino, Italy
}\\[0.5cm]
{\em${}^b$ Particle Theory and Cosmology Group, \\ 
Department of Physics, Graduate School of Science, \\
Tohoku University\\
Aoba-ku, Sendai 980-8578, Japan}\\[0.5cm]
{\em${}^c$ Maxwell Institute for Mathematical Sciences\\
Department of Mathematics,
Heriot--Watt University\\
Edinburgh EH14 4AS, U.K.}\\[0.5cm]
{Email: {\ttfamily deser@to.infn.it, marcandre.heller@hotmail.com, c.saemann@hw.ac.uk}}
\end{center}
\vskip 1.0cm
\begin{center}
{\bf Abstract}
\end{center}
\begin{quote}
We continue our exploration of local Double Field Theory (DFT) in terms of symplectic graded manifolds carrying compatible derivations and study the case of heterotic DFT. We start by developing in detail the differential graded manifold that captures heterotic Generalized Geometry which leads to new observations on the generalized metric and its twists. We then give a symplectic pre-N$Q$-manifold that captures the symmetries and the geometry of local heterotic DFT. We derive a weakened form of the section condition, which arises algebraically from consistency of the symmetry Lie 2-algebra and its action on extended tensors. We also give appropriate notions of twists---which are required for global formulations---and of the torsion and Riemann tensors. Finally, we show how the observed $\alpha'$-corrections are interpreted naturally in our framework.
\end{quote}

\end{titlepage}

\tableofcontents

\section{Introduction and results}

Double Field Theory (DFT) is the attempt of finding a T-dual invariant formulation of the low-energy sector of string theory. After initial work in the 90ies~\cite{Tseytlin:1990va,Siegel:1993th,Siegel:1993xq}, the field received more attention after the papers~\cite{Hull:2004in,Hull:2009sg} and in particular~\cite{Hull:2009mi} appeared. While many of the physical aspects of DFT are well studied and mostly understood by now, the underlying mathematical structures are much less explored (see e.g.~\cite{Vaisman:2012ke,Vaisman:2012px,Deser:2014mxa,Freidel:2017yuv} for rare exceptions).

DFT is clearly an extension of gravity coupled to a 2-form gauge potential (the Kalb--Ramond or $B$-field), whose underlying geometry is famously described by Generalized Geometry~\cite{Hitchin:2004ut,Hitchin:2005in,Gualtieri:2003dx}. Generalized Geometry and its symmetry structure is encoded in a Courant algebroid, which for our purposes is most transparently described in terms of symplectic differential graded (dg) manifolds or symplectic N$Q$-manifolds~\cite{Roytenberg:0203110}. For example, the Courant bracket is simply part of a categorified Lie algebra, which is obtained from a simple derived bracket construction in the language of dg-manifolds. It is, in fact, the symmetry Lie 2-algebra which consists of the semidirect product of diffeomorphisms on the base manifold and 1-form gauge transformations of the gerbe with the same topological class as that of the Courant algebroid. Moreover, twists, T-dualities and even the structure of the generalized metric can be directly obtained using canonical transformations of the symplectic dg-manifold. Similarly, one readily constructs expressions for torsion and Riemann tensors. Finally, a global Courant algebroid is readily constructed in terms of local data contained in the dg-manifold picture.

All this suggests that also DFT should have a useful description in terms of symplectic graded manifolds, at least locally. This was the premise of the paper~\cite{Deser:2014mxa} and in particular~\cite{Deser:2016qkw}. In the latter work, the general framework was developed and applied to the simplest form of local DFT, in the following called ``type II DFT''. Among other things, the symplectic graded manifold picture yields directly the full relevant symmetry Lie 2-algebra, a resulting algebraic section condition (which is a slight weakening of the usual one) and explicit expressions for torsion and Riemann tensors. Mathematically, the key ingredient in our constructions was a relaxation of the notion of a symplectic dg-manifold to symplectic graded manifolds with a degree~1 vector field $Q$ not necessarily satisfying $Q^2=0$. Essentially, the same framework was subsequently used in~\cite{Heller:2016abk} to give a unified approach to non-geometric fluxes and T-duality in terms of canonical transformations. Furthermore, the symmetry Lie 2-algebra obtained in this manner was also identified in~\cite{Hohm:2017pnh} as the appropriate one underlying DFT.

Our long-term goal is the extension of our framework to exceptional field theory (EFT). As a stepping stone, we explore in this paper the case of the bosonic part of heterotic Double Field Theory, which already contains some of the important new features we expect to see in EFT. We will simply refer to this form of DFT as heterotic DFT.

Heterotic DFT has been discussed in various forms, starting from the initial paper~\cite{Hohm:2011ex}. It features in particular a number of $\alpha'$-corrections as, for example, to the usual pairing of generalized vectors and the C- and D-brackets of type II DFT~\cite{Hohm:2013jaa,Hohm:2014xsa,Hohm:2014eba}. These corrections only appear if the generalized vector fields of heterotic DFT are identified with ordinary vector fields and differential 1-forms, just as in type II DFT. Since heterotic string theory (as well as heterotic supergravity) also comes with a non-abelian gauge algebra, there is an enlarged set of symmetries, which should be reflected in a larger set of generalized vector fields. In particular, one should incorporate further sections that describe the ordinary gauge transformations. In such a formulation, heterotic DFT becomes more transparent~\cite{Blumenhagen:2014iua} and, as we shall explain below, the $\alpha'$-corrections find a  very natural interpretation. 

Our discussion starts in section~\ref{sec:GG} by a reformulation of heterotic Generalized Geometry in terms of symplectic differential graded or symplectic N$Q$-manifolds. The underlying transitive Courant algebroid was given before in more conventional geometric language in~\cite{Garcia-Fernandez:2013gja, Baraglia:2013wua}. We give the corresponding N$Q$-manifold perspective, from which we directly derive the associated Lie 2-algebra from a derived bracket construction. This categorified Lie algebra appears in the form of a 2-term $L_\infty$-algebra and describes the symmetries of heterotic supergravity: a semidirect product of diffeomorphisms on the base manifold and gauge transformations, which consist of ordinary gauge transformations of a non-abelian 1-form gauge potential and higher gauge transformations of an abelian 2-form gauge potential. The heterotic Dorfman and Courant brackets are part of this Lie 2-algebra structure. 

Recall that in ordinary Generalized Geometry, the exact Courant algebroid describes the symmetries of an abelian gerbe, on which the Kalb--Ramond field is a part of the connective structure~\cite{Deser:2016qkw}. We point out that the transitive Courant algebroid describes the symmetries of a non-abelian gerbe (or categorified principal bundle) whose categorified structure group is the string 2-group. Such structures were fully developed only in~\cite{Sati:2009ic} and they play an important role in M-theory, see e.g.~\cite{Fiorenza:2012tb,Saemann:2017rjm}.

We proceed to interpret twists of the transitive Courant algebroid by topological data as canonical transformations of our N$Q$-manifold. This is a specialization of a result of~\cite{Roytenberg:0112152} to the case at hand. We use this interpretation to rederive the curvature expressions for twisted string structures together with the relevant Bianchi identities in a very direct fashion. A new observation is that also the generalized metric for heterotic Generalized Geometry can be obtained from the trivial one by a canonical transformation, thanks to our formalism. Finally, the N$Q$-manifold picture allows us to write down suitable torsion and Riemann tensors that yield a compact description of heterotic supergravity by using the formulas developed in~\cite{Deser:2016qkw} for type II DFT.

Section~\ref{sec:hDFT} contains the main point of our paper: a similar treatment of heterotic DFT, building on the formalism developed in~\cite{Deser:2016qkw}. Analogously to the case of type II DFT, we have to work here with symplectic pre-N$Q$-manifolds, in which the condition $Q^2=0$ is lifted. The symplectic pre-N$Q$-manifold relevant to heterotic DFT is readily written down from the relevant data. As expected, it extends the symplectic N$Q$-manifold of heterotic Generalized Geometry as well as the symplectic pre-N$Q$-manifold of type II DFT. The derived bracket construction reproduces the C- and D-brackets of heterotic DFT. Moreover, the condition that this construction yields a Lie 2-algebra of symmetries acting nicely on generalized tensor fields yields a weakened form of the strong section condition usually given in heterotic DFT, cf.~\cite{Hohm:2011ex}. Also, the notion of twist by topological data extends to heterotic DFT, which is crucial for the construction of global formulations of this theory. In particular, we derive twist data and the generalized metric from canonical transformations, just as in the cases of heterotic Generalized Geometry and partially extending the case of type II DFT~\cite{Heller:2016abk}. Similarly, we use the same formulas as in type II DFT and heterotic Generalized Geometry to define appropriate torsion and Riemann tensors. Once more, we stress that all of these constructions emerge in a straightforward and natural way from the symplectic pre-N$Q$-manifold from which we started.

We conclude in section~\ref{sec:alphaprime} with a discussion of the issue of $\alpha'$-corrections in heterotic DFT. As stated initially, these corrections arise if heterotic DFT is described using the same generalized vector fields as in type II DFT. In~\cite{Deser:2014wva}, it was observed that these $\alpha'$-corrections of the pairing and the C- and D-brackets can be traced back to a deformation of the Poisson bracket, which was further interpreted as a degree-violating star-product. 

We extend this interpretation via a deformed Poisson bracket to the full Lie 2-algebra for both heterotic Generalized Geometry as well as heterotic DFT. We argue, however, that a more direct explanation than the degree-violating star product is obtained from geometric considerations. In the case of Generalized Geometry, this observation is due to~\cite{Coimbra:2014qaa}: Diffeomorphisms induce local transformations of the frame bundle and each generalized vector field $X$ therefore comes with a local transformation $\dpar_\mu X^\nu$ of the frame bundle. In the ``uncorrected picture'', by which we mean the extended formulation of heterotic DFT which we used in section~\ref{sec:hDFT}, sections of the frame bundle are already included. To connect both formulations, we should translate ordinary extended vector fields $X$ in the $\alpha'$-corrected picture to extended vector fields $X+\dpar_\mu X^\nu$ with $\dpar_\mu X^\nu$ a section of the frame bundle in the uncorrected picture. The usual derivative counting shows that either the term $\dpar_\mu X^\nu$ should come with a factor of $\sqrt{\alpha'}$ or, more appropriately, the inner product on the frame bundle should come with a factor of $\alpha'$. In any case, we indeed recover the deformation of the Poisson bracket which induces all other expected $\alpha'$-corrections.

This interpretation also generalizes to the case of DFT and furthermore answers a question from~\cite{Bedoya:2014pma}: There, it was observed that certain terms should be added to generalized vector fields to make the transition between the $\alpha'$-corrected and the uncorrected formulation of heterotic DFT. It is a generalization of the interpretation of~\cite{Coimbra:2014qaa} to Double Field Theory which is responsible for this. Interestingly, the resulting C- and D-brackets contain two further terms in the $\alpha'$-corrections of DFT which, however, vanish in the usual formulations of heterotic DFT due to the strong section condition. Our weakened form of the strong section condition, however, does not imply their vanishing.

Altogether, we conclude that we found a second successful application of the language of symplectic pre-N$Q$-manifolds to the description of the geometric structures underlying DFT. We are therefore confident that it is suitable for an extension to EFT. We plan to report on progress in this direction in an upcoming publication.

\section{Generalized Geometry}\label{sec:GG}

\subsection{The heterotic string and twisted string structures}

Heterotic string theory~\cite{Gross:1985fr,Gross:1985rr} is a chirally asymmetric combination of the ten-dimensional type II superstring and the 26-dimensional bosonic string, where the former and the latter constitute the left- and right-moving sectors of the heterotic string, respectively. To obtain a consistent supersymmetric ten-dimensional theory which is Lorentz invariant, the additional 16~dimensions of the bosonic string are compactified on a particular 16-dimensional torus, whose properties enhance the local symmetry group from $\sU(1)^{16}$ to $\sSO(32)$ or $\sE_8\times \sE_8$. Although heterotic string theory is a theory of closed strings and D-branes are absent, it thus gives rise to a gauge group, and the spectrum contains massless modes corresponding to non-abelian gauge potentials besides the metric, the dilaton and the Kalb--Ramond field.

It has been known for a long time that consistently coupling a gauge potential to ten-dimensional supergravity requires adding a Chern--Simons term to the definition of the 3-form curvature $H$ of the 2-form potential $B$~\cite{Bergshoeff:1981um,Chapline:1982ww}. This also affects Bianchi identities and the gauge structure. Altogether, the bosonic part of heterotic supergravity (or the bosonic massless sector of the heterotic string) is given by the metric $g$, the dilaton $\phi$, a spin connection 1-form $\omega$ taking values in $\aspin(1,9)$, a gauge potential 1-form $A$ taking values in $\aso(32)$ or $\ae_8\times \ae_8$ and the Kalb--Ramond $B$-field, which is a 2-form taking values in $\au(1)$. We shall denote the Killing forms on the underlying Lie algebras by $(-,-)$. The relevant curvatures are defined as
\begin{subequations}\label{eq:twisted_string_structures}
\begin{equation}\label{eq:gg_fs}
 F_\omega=\dd \omega+\tfrac12 [\omega,\omega]~,~~~F_A=\dd A+\tfrac12[A,A]~,~~~H=\dd B-{\rm cs}(\omega)+{\rm cs}(A)~,
\end{equation}
where ${\rm cs}(A)=(A,\dd A)+\tfrac13(A,[A,A])$ is the usual Chern--Simons form. The infinitesimal gauge transformations read as\footnote{Note that in~\cite{Hohm:2014eba}, a different set of infinitesimal gauge transformations was found, which is equivalent to the more conventional one given here after a shift of the 1-form gauge parameter $\Lambda$.}
\begin{equation}
\begin{aligned}
 \delta \omega &= \dd \lambda + [\omega,\lambda]~,\\
 \delta A &= \dd\alpha + [A,\alpha]~,\\
 \delta B &= \dd \Lambda +(\lambda,\dd \omega)-(\alpha,\dd A)~,
\end{aligned}
\end{equation}
where $\lambda$, $\alpha$ and $\Lambda$ parameterize gauge transformations of $\omega$, $A$ and $B$. Note that these gauge transformations leave $H$ invariant. The relevant Bianchi identities read as
\begin{equation}\label{eq:gg_Bianchi}
 \dd F_\omega+[\omega,F_\omega]=0~,~~~\dd F_A+ [A,F_A]=0~,~~~\dd H=(F_\omega,F_\omega)-(F_A,F_A)~,
\end{equation}
the last of which is the {\em Green--Schwarz anomaly cancellation condition}, which guarantees that the gauge and the gravitational anomalies cancel each other~\cite{Green:1984sg}.
\end{subequations}

The structure above has only found a comprehensive mathematical interpretation in~\cite{Sati:2009ic}, see also~\cite{Sati:2008eg}. Recall that the $B$-field of bosonic and type II string theory is part of the connective structure of an abelian gerbe~\cite{Gawedzki:1987ak,Freed:1999vc}. It turns out that the fields $\omega$, $A$ and $B$ form part of the connective structure of a non-abelian gerbe with structure group $\sString(1,9)$, which is twisted by $\sE_8\times \sE_8$ and~\eqref{eq:twisted_string_structures} form the (local) description of such a connective structure. These {\em twisted string structures} feature also crucially in effective descriptions of M5-branes, see e.g.~\cite{Fiorenza:2012tb,Saemann:2017rjm}. 

A full description of twisted string structures would take us too far away from the aim of our discussion and we merely refer to the original paper~\cite{Sati:2009ic} as well as~\cite{Saemann:2017rjm} for details. For our purposes, it suffices to keep in mind formulas~\eqref{eq:twisted_string_structures} as well as the following summary of the string Lie 2-algebra.

The string group $\sString(n)$ is defined as an element of the Whitehead tower of $\sO(n)$, namely as the 3-connected cover of $\sSpin(n)$. This definition fixes $\sString(n)$ only up to a large class of equivalences, and there exist various models of it. Particularly convenient models are given in terms of categorified Lie groups or Lie 2-groups. These Lie 2-group models can be Lie differentiated, leading to Lie 2-algebra models\footnote{see the appendix for relevant definitions} of $\astring(n)$. A particularly simple model of $\astring(n)$ is given by 
\begin{subequations}\label{eq:string_lie_2_algebra}
\begin{equation}
 \astring(n)\ =\ \big(~\FR~\xrightarrow{~0~} \aspin(n)~\big)~,
\end{equation}
with non-trivial products
\begin{equation}
\begin{aligned}
 \mu_2&:\aspin(n)\wedge \aspin(n)\rightarrow \aspin(n)~,~~~&\mu_2(x_1,x_2)&=[x_1,x_2]~,\\
 \mu_3&:\aspin(n)\wedge \aspin(n)\wedge \aspin(n)\rightarrow \FR~,~~~&\mu_3(x_1,x_2,x_3)&=(x_1,[x_2,x_3])~.
\end{aligned}
\end{equation}
\end{subequations}
Here, $(-,-)$ is again the appropriately normalized Killing form on $\aspin(n)$. 

Clearly, this is the special case of a general class of Lie 2-algebras which were first considered in~\cite{Baez:2003aa}. Historically, the integration of $\astring(n)$ to a string group model came first~\cite{Henriques:2006aa,Baez:2005sn}; a Lie differentiation leading to $\astring(n)$ was presented in~\cite{Demessie:2016ieh}.

For our purposes, the local definition~\eqref{eq:twisted_string_structures} of a connective structure on a principal 2-bundle with structure group $\sString(n)$ is sufficient. We merely note in addition that the action of heterotic supergravity on some $D$-dimensional manifold $M$ reads as~\cite{Gross:1985rr}
\begin{equation}
\label{hetaction}
 S=\int \dd^Dx \sqrt{g}~\de^{-2\phi}\big(R-\tfrac{1}{12}H^2-\tfrac{1}{4}\tr(F_\omega^2)+\tfrac{1}{4}\tr(F_A^2)\big)~,
\end{equation}
where $g$ is the metric and $R$ the Ricci scalar.

\subsection{Symmetries of string structures: Heterotic Courant algebroids}\label{ssec:het_Courant_algebroids}

We start by recalling the case of type II supergravity on some manifold $M$, where the Kalb--Ramond field $B$ is part of a connective structure on an abelian gerbe $\CCG$ with curvature $H=\dd B$ over $M$. The infinitesimal symmetries of this gerbe are captured by the exact Courant algebroid $E$ which fits into the sequence $T^*M\rightarrow E\rightarrow TM$ and which has \v Severa class $H$. Since all vector bundles are real, this sequence splits and $E\cong TM\oplus T^*M$. Sections of $E$ are therefore given by a vector field and a 1-form and parameterize infinitesimal diffeomorphisms as well as gauge transformations of $B$. As shown in~\cite{Roytenberg:0203110}, the Courant algebroid is most appropriately regarded as a symplectic Lie 2-algebroid, cf.~also the appendix. Recall that any symplectic Lie $n$-algebroid comes with an associated Lie $n$-algebra, cf.\ the appendix. The associated Lie 2-algebra of $E$ is now the $L_\infty$-algebra of infinitesimal symmetries of the gerbe $\CCG$.

This picture readily generalizes to the case of heterotic supergravity. Here, we also have a principal bundle $P$ over $M$ which is the combination of the gauge bundle with gauge group $\sSO(32)$ or $\sE_8\times \sE_8$ as well as the frame bundle underlying the spin connection. Locally, the relevant Courant algebroid is then
\begin{equation}\label{eq:transitive_CA}
 E\cong TM\oplus {\rm ad} P\oplus T^*M~,
\end{equation}
where ${\rm ad}P$ is the trivial Lie algebra bundle $M\times \frg\rightarrow M$ with
\begin{equation}
 \frg=\aspin(1,9)\oplus \ae(8)\oplus \ae(8)~.
\end{equation}
We endow $\frg$ with an indefinite Killing form, which is positive and negative definite on the summands $\aspin(1,9)$ and $\ae(8)\oplus \ae(8)$, respectively.

The transitive Courant algebroid~\eqref{eq:transitive_CA} was first mentioned in~\cite{Severa:1998ac}. The connection to the Green--Schwarz anomaly was discussed in~\cite{Garcia-Fernandez:2013gja,Garcia-Fernandez:2016ofz} and in~\cite{Baraglia:2013wua}, the full global picture was worked out and its role in topological T-duality was explained. We also follow the nomenclature in~\cite{Baraglia:2013wua} and call $E$ the {\em heterotic Courant algebroid}.

For our purposes, it will be useful to describe $E$ as a symplectic N$Q$-manifold. We start from the N-manifold
\begin{equation}
 \CW_2(M,P)=T^*[2]T[1]M\oplus {\rm ad}P^*[1]
\end{equation}
for some contractible space $M$ of dimension $D$. We introduce coordinates $x^\mu$, $\mu=1,\dots,D$, on the base of $\CW_2(M,P)$ as well as coordinates $\xi^\mu,\zeta_\mu,p_\mu$ and $\tau_\alpha$, $\alpha=1,\dots,\dim \frg$, in the fibers. The degrees of the various coordinates are as follows:
\begin{center}
\begin{tabular}{|l|c|c|c|c|c|}
 \hline
 coordinate & $x^\mu$ & $\xi^\mu$ & $\zeta_\mu$ & $p_\mu$ & $\tau_\alpha$\\
 \hline
 degree & 0 & 1 & 1 & 2 & 1 \\
 \hline
\end{tabular}
\end{center}
The N$Q$-manifold $\CW_2(M,P)$ comes with a natural symplectic form,
\begin{equation}
 \omega=\dd x^\mu\wedge \dd p_\mu+\dd \xi^\mu\wedge \dd \zeta_\mu+\tfrac12 \kappa^{\alpha\beta}\dd \tau_\alpha\wedge \dd \tau_\beta~,
\end{equation}
where $\kappa^{\alpha\beta}$ is the inverse of the Killing form $\kappa_{\alpha\beta}$. The symplectic form induces the Poisson bracket
\begin{equation}
 \{f,g\}:=f\overleftarrow{\der{p_\mu}}\overrightarrow{\der{x^\mu}} g-f\overleftarrow{\der{x^\mu}}\overrightarrow{\der{p_\mu}} g-f\overleftarrow{\der{\zeta_\mu}}\overrightarrow{\der{\xi^\mu}} g-f\overleftarrow{\der{\xi^\mu}}\overrightarrow{\der{\zeta_\mu}} g-f\overleftarrow{\der{\tau_\alpha}}\kappa_{\alpha\beta}\overrightarrow{\der{\tau_\beta}} g~,
\end{equation}
where $f,g$ are functions on $\CW_2(M,P)$. Let $f^\alpha_{\beta\gamma}$ be the structure constants of $\frg$ and define $f^{\alpha\beta\gamma}=f^\alpha_{\delta\eps}\kappa^{\beta\delta}\kappa^{\gamma\eps}$ as the structure constants with all indices raised by the Killing form. Then the function
\begin{equation}
 \CQ=\xi^\mu p_\mu-\tfrac{1}{3!}f^{\alpha\beta\gamma}\tau_\alpha\tau_\beta\tau_\gamma
\end{equation}
is the Hamiltonian function for a homological vector field $Q$ since the Jacobi identity implies $\{\CQ,\CQ\}=0$.

More generally, we consider the twisted version
\begin{equation}
 \CQ=\xi^\mu p_\mu-\tfrac{1}{3!}f^{\alpha\beta\gamma}\tau_\alpha\tau_\beta\tau_\gamma+\tfrac12 A^\alpha_\mu\xi^\mu \kappa_{\alpha\beta}f^{\beta\gamma\delta}\tau_\gamma\tau_\delta-\tfrac{1}{2!}F^\alpha_{\mu\nu}\xi^\mu\xi^\nu\tau_\alpha-\tfrac{1}{3!}H_{\mu\nu\kappa}\xi^\mu\xi^\nu\xi^\kappa~,
\end{equation}
for which $\{\CQ,\CQ\}=0$ if in addition to the Jacobi identity, we have\footnote{When comparing this formula with~\eqref{eq:twisted_string_structures}, recall that we combined $\aspin(1,9)\oplus \ae(8)\oplus \ae(8)$ into $\frg$ with indefinite Killing form and accordingly $\omega+A$ into $A$.}
\begin{equation}\label{eq:GG_twist_condition}
 \dd A+\tfrac12 [A,A]=F~,~~~\nabla F=0\eand \dd H=(F,F)~.
\end{equation}
That is, the generalized \v Severa class of the transitive Courant algebroid $\CW_2(M,P)$ is given by the pair $(F,H)$, which also characterizes a twisted string structure as we saw above.

The homological vector field $Q=\{\CQ,-\}$ is readily computed, and we obtain
\begin{equation}\label{eq:Q_heterotic_GG}
\begin{aligned}
 Q&=\xi^\mu\der{x^\mu}
  +p_\mu\der{\zeta_\mu}
  -\frac{1}{2} \tau_\alpha\tau_\beta f^{\alpha\beta\gamma}\kappa_{\gamma\delta}\der{\tau^\delta}\\
  &~~~-\frac{1}{2}\tau_\alpha\tau_\beta A^\gamma_\mu \kappa_{\gamma\delta}f^{\alpha\beta\delta}\der{\zeta_\mu}
  +\frac{1}{2}\xi^\mu\tau_\alpha\tau_\beta\der{x^\nu}A^\gamma_\mu\kappa_{\gamma\delta}f^{\alpha\beta\gamma}\der{p_\nu}
  +\xi^\mu\tau_\alpha A^\beta_\mu f^\alpha_{\beta\gamma}\der{\tau_\gamma}\\
  &~~~-\frac{1}{2}\xi^\mu\xi^\nu\tau_\alpha\der{x^\kappa}F^\alpha_{\mu\nu}\der{p_\kappa}
  -\xi^\mu\tau_\alpha F^\alpha_{\mu\nu}\der{\zeta_\nu}
  +\frac{1}{2}\xi^\mu\xi^\nu F^\alpha_{\mu\nu}\kappa_{\alpha\beta}\der{\tau_\beta}\\
  &~~~+\frac{1}{3!}\xi^\mu\xi^\nu\xi^\kappa\der{x^\lambda}H_{\mu\nu\kappa}\der{p_\lambda}
  -\frac{1}{2} \xi^\mu\xi^\nu H_{\mu\nu\kappa}\der{\zeta_\kappa}~.
\end{aligned}
\end{equation}
This completes the description of the symplectic N$Q$-manifold $\CW_2(M,P)$. 

The Dorfman bracket of the heterotic Courant algebroid as given in~\cite{Baraglia:2013wua} can now be obtained from a natural derived bracket construction:
\begin{equation}\label{eq:Dorfman_bracket}
\begin{aligned}
 \nu_2(X_1+\lambda_1+\alpha_1,X_2+\lambda_2+\alpha_2)&=\{Q(X_1+\lambda_1+\alpha_1),X_2+\lambda_2+\alpha_2\}\\
 &=[X_1,X_2]+\CL_{X_1}\alpha_2-\iota_{X_2}\dd \alpha_1\\
 &~~~~+\nabla_{X_1}\lambda_2-\nabla_{X_2}\lambda_1+(\nabla \lambda_1,\lambda_2)+[\lambda_1,\lambda_2]\\
 &~~~~+(\iota_{X_1}F,\lambda_2)-(\iota_{X_2}F,\lambda_1)-F(X_1,X_2)-\iota_{X_1}\iota_{X_2}H~.
\end{aligned}
\end{equation}
Here, $[-,-]$ and $(-,-)$ denote Lie bracket and Killing form on $\frg$, respectively. Moreover, $X_i=X^\mu_i\zeta_\mu$, $\lambda_i=\lambda^\alpha_i\tau_\alpha$ and $\alpha_i=\alpha_{i\mu}\xi^\mu$ and we regard $X_i+\lambda_i+\alpha_i$ as a section of $E$.

The local infinitesimal symmetries of a twisted string structure are now captured by the associated Lie 2-algebra\footnote{See the appendix for details.} of $\CW_2(M,P)$. This Lie 2-algebra has underlying graded vector space
\begin{equation}
 \sL=~\big(~\CC^\infty(M)~\rightarrow~\Gamma(E)~\big)
\end{equation}
and the higher brackets are given by the totally antisymmetric multilinear maps
\begin{equation}
\begin{aligned}
 \mu_1(f+X+\lambda+\alpha)&=\dd f= \xi^\mu \dpar_\mu f~,\\
 \mu_2(X_1+\lambda_1+\alpha_1,X_2+\lambda_2+\alpha_2)&=[X_1,X_2]+\CL_{X_1}\alpha_2-\CL_{X_2}\alpha_1-\tfrac12\dd(\iota_{X_1}\alpha_2-\iota_{X_2}\alpha_1)\\
 &~~~~+\nabla_{X_1}\lambda_2-\nabla_{X_2}\lambda_1+\tfrac12(\nabla \lambda_1,\lambda_2)-\tfrac12(\lambda_1,\nabla \lambda_2)\\
 &~~~~+[\lambda_1,\lambda_2]+(\iota_{X_1}F,\lambda_2)-(\iota_{X_2}F,\lambda_1)-F(X_1,X_2)\\
 &~~~~-\iota_{X_1}\iota_{X_2}H~,\\
 \mu_2(X_1+\lambda_1+\alpha_1,f)&=\tfrac12 X_1 f\\
 \mu_3(X_1+\dots,X_2+\dots,X_3+\dots)&=\tfrac{1}{3!}\big(2\iota_{X_1}\iota_{X_2}\dd \alpha^3+\iota_{X_3}\dd \iota_{X_1}\alpha^2+\iota_{X_1}(\nabla \lambda_2,\lambda_3)+\\
 &~~~~~~~+\iota_{X_1}\iota_{X_2}(F,\lambda_3)+{\rm perm.}\big)+\\
 &~~~~+(\lambda_1,[\lambda_2,\lambda_3])-\iota_{X_1}\iota_{X_2}\iota_{X_3}H
\end{aligned}
\end{equation}
for $f\in \CC^\infty(M)$ and $X_i+\lambda_i+\alpha_i\in \Gamma(E)$. We note that for $X_i=\alpha_i=0$ and $f$ and $\lambda_i$ constant, this Lie 2-algebra reduces to the string Lie 2-algebra~\eqref{eq:string_lie_2_algebra} with $\aspin(n)$ replaced by $\frg$.

The symmetries of the string structure now act on various tensors, which are particular elements of the free tensor algebra $\CT(\CC^\infty(\CM))$ of $\CC^\infty(\CM)$ over $\CC^\infty_0(\CM)$. The action of infinitesimal symmetries $X_1+\lambda_1+\alpha_1\in \Gamma(E)$ on generalized vectors $X_2+\lambda_2+\alpha_2$ is captured by the Dorfman bracket~\eqref{eq:Dorfman_bracket}. More generally, we define an action on $t\in \CT(\CC^\infty(\CM))$ by
\begin{equation}
 \hat \CL_X t=\nu_2(X,t):=\{Q X,t\}~,
\end{equation}
where we extended the Poisson bracket in the second slot via
\begin{equation}\label{eq:ext_Poisson}
 \{f,g\otimes h\}:=\{f,g\}\otimes h+(-1)^{(2-|f|)|g|}g\otimes \{f,h\}~,
\end{equation}
as done in~\cite{Deser:2016qkw}. For example, we can compute the action of an infinitesimal symmetry on the generalized metric $\CH$:
\begin{equation}
 \hat \CL_X \CH=\nu_2(X,\CH)=\{Q X,\CH\}~,
\end{equation}
where $\CH$ is the generalized metric
\begin{equation}\label{eq:xiM}
 \CH_{MN}~\xi^M\otimes \xi^N:=\CH_{\mu\nu}~\xi^\mu\otimes \xi^\nu+\CH_\mu{}^\nu~\xi^\mu\otimes \zeta_\nu+\CH^\mu{}_\nu~\zeta_\mu\otimes\xi^\nu+\CH^{\mu\nu}~\zeta_\mu\otimes \zeta_\nu~,
\end{equation}
which we shall discuss in more detail below.

\subsection{Topological data from twists}

Let us return to the trivial heterotic Courant algebroid $\CW_2(M,M\times \frg)$ with $A=B=0$. Turning on the potentials $A$ and $B$ for the topological data $F$ and $H$ can be interpreted as the result of a canonical transformation, as we shall show in the following.  This is a specialization of a general result of~\cite{Roytenberg:0112152}.

The trivial heterotic Courant algebroid has homological vector field given by the Hamiltonian
\begin{equation}
 \CQ_0=\xi^\mu p_\mu-\tfrac{1}{3!}f^{\alpha\beta\gamma}\tau_\alpha\tau_\beta\tau_\gamma~.
\end{equation}
On a function $f$ on $\CW_2(M,P)$ (and, by extension of the Poisson bracket~\eqref{eq:ext_Poisson}, also on tensor products of these), we can define canonical transformations given by the flow of the Hamiltonian vector field $X_\phi$ of a function $\phi\in \CC^\infty(\CW_2(M,P))$ according to
\begin{equation}
 \de^{\phi}\acton f:=\de^{X_\phi} f=\de^{\{\phi,-\}}f=f+\{\phi,f\}+\tfrac{1}{2} \{\phi,\{\phi,f\}\}+\dots
\end{equation}
Clearly, $\phi$ has to be of degree~2 in order to preserve the degree of $f$. The most general such Hamiltonian on $\CW_2(M,P)$ reads as 
\begin{equation}
 \phi=\tfrac12 B_{\mu\nu}\xi^\mu\xi^\nu+\tilde B_\mu{}^\nu \xi^\mu\zeta_\nu+\tfrac12 \beta^{\mu\nu}\zeta_\mu\zeta_\nu+A_\mu^\alpha\xi^\mu\tau_\alpha+\tilde A^{\mu\alpha}\zeta_\mu\tau_\alpha+a^{\alpha\beta}\tau_\alpha\tau_\beta+t^\mu p_\mu~,
\end{equation}
where each of the coefficients is a function on $M$ and prefactors are inserted for later convenience. We immediately note that $t^\mu p_\mu$ amounts to a diffeomorphism on $M$ and that $\tilde B_\mu{}^\nu$ and $a^{\alpha\beta}$ are merely rotations of the coordinate systems on the fibers of the generalized tangent bundle $TM\oplus \ad P\oplus T^*M$. Since these transformations are trivial, we put the corresponding coefficients to zero.

The remaining twist data is packaged into two pairs, $(A,B)$ and\footnote{We follow the literature and use the more common notation $\beta$ instead of the more logical $\tilde B$.} $(\tilde A,\beta)$ to guarantee that the power series $\de^\phi$ terminates and therefore clearly converges. We also apply the corresponding canonical transformations step-by-step, for clarity. We observe that 
\begin{equation}\label{eq:twist_gg}
\begin{aligned}
 &\CQ=\de^{A^\alpha_\mu \xi^\mu\tau_\alpha}\acton\left(\de^{\tfrac12 B_{\mu\nu}\xi^\mu\xi^\nu}\acton\CQ_0\right)\\
 &\hspace{0.5cm}=\xi^\mu p_\mu-\tfrac{1}{3!}f^{\alpha\beta\gamma}\tau_\alpha\tau_\beta\tau_\gamma+\tfrac12 A^\alpha_\mu\xi^\mu \kappa_{\alpha\beta}f^{\beta\gamma\delta}\tau_\gamma\tau_\delta-\tfrac{1}{2!}F^\alpha_{\mu\nu}\xi^\mu\xi^\nu\tau_\alpha-\tfrac{1}{3!}H_{\mu\nu\kappa}\xi^\mu\xi^\nu\xi^\kappa~,
\end{aligned}
\end{equation}
which is the Hamiltonian on $\CW_2(M,P)$ for nontrivial $F$ and $H$. In this way, we readily obtain the correct definition of the field strengths for a string structure~\eqref{eq:gg_fs}, even if we had not been aware of it. Moreover, the condition $\{\CQ,\CQ\}=0$ yields the relevant Bianchi identities~\eqref{eq:gg_Bianchi} in a straightforward manner.

We also find 
\begin{equation}
\begin{aligned}
 \tilde \CQ&=\de^{\tilde A^{\mu\alpha} \zeta_\mu\tau_\alpha}\acton\left(\de^{\tfrac12 \beta^{\mu\nu}\zeta_\mu\zeta_\nu}\acton\CQ_0\right)\\
 &=\xi^\mu p_\mu-\tfrac{1}{3!}f^{\alpha\beta\gamma}\tau_\alpha\tau_\beta\tau_\gamma
 -\tfrac12 \xi^\mu\dpar_\mu \beta^{\nu\kappa}\zeta_\nu\zeta_\kappa
 +\tfrac12 \beta^{\mu\nu}\zeta_\mu p_\nu
 -\tfrac14 (\dpar_\nu \beta^{\kappa\lambda})\beta^{\mu\nu}\zeta_\mu\zeta_\kappa\zeta_\lambda\\
 &~~-\tilde A^{\mu\alpha}p_\mu\tau_\alpha 
 -\dpar_\mu \tilde A^{\nu\alpha}\xi^\mu\zeta_\nu\tau_\alpha
 -\tfrac12 \tilde A^{\mu\alpha}\zeta_\mu\kappa_{\alpha\beta}f^{\beta\gamma\delta}\tau_\gamma\tau_\delta
 -\tfrac12 \tilde A^{\mu\alpha}\tau_\alpha\dpar_\mu\beta^{\nu\kappa}\zeta_\nu\zeta_\kappa\\
 &~~-\tfrac12 (\dpar_\nu \tilde A^{\kappa\alpha})\beta^{\mu\nu}\zeta_\kappa\tau_\alpha\zeta_\mu\\
 &~~+\frac{1}{2}\Big(
 -\tilde A^{\mu\alpha}\zeta_\mu\kappa_{\alpha\beta}\tilde A^{\nu\beta}p_\nu
 +2(\dpar_\mu \tilde A^{\nu\alpha})\zeta_\nu\tau_\alpha \tilde A^{\mu\beta}\tau_\beta
 -\tilde A^{\mu\alpha}\zeta_\mu\kappa_{\alpha\beta}\dpar_\nu \tilde A^{\kappa\beta}\zeta_\kappa\xi^\nu\Big.\\
 &~~~~~~~~~~~\Big.
 -\tilde A^{\mu\alpha}\zeta_\mu \tilde A^{\nu\beta}\zeta_\nu\kappa_{\alpha\gamma}\kappa_{\beta\delta}f^{\gamma\delta\eps}\tau_\eps
 -\tfrac12 \tilde A^{\mu\alpha}\zeta_\mu\kappa_{\alpha\beta}\tilde A^{\nu\beta}\dpar_\nu \beta^{\kappa\lambda}\zeta_\kappa\zeta_\lambda\Big.\\
 &~~~~~~~~~~~\Big.+\tfrac12 \tilde A^{\mu\alpha}\zeta_\mu\kappa_{\alpha\beta}(\dpar_\kappa \tilde A^{\lambda \beta})\beta^{\nu\kappa}\zeta_\lambda\zeta_\nu\Big)\\
 &~~+\frac{1}{3!}\Big(
 (\dpar_\mu \tilde A^{\nu\alpha})\zeta_\nu\tau_\alpha\tilde A^{\kappa\beta}\zeta_\kappa\kappa_{\beta\gamma}\tilde A^{\mu\gamma}
 +2\tilde A^{\mu\alpha}\zeta_\mu\kappa_{\alpha\beta}(\dpar_\nu\tilde A^{\kappa\beta})\zeta_\kappa \tilde A^{\nu\gamma}\tau_\gamma\Big.\\
 &~~~~~~~~~~~\Big.
 -2\tilde A^{\mu\alpha}\zeta_\mu\kappa_{\alpha\beta}(\dpar_\nu\tilde A^{\kappa\gamma})\zeta_\kappa\tau_\gamma \tilde A^{\nu\beta}
 -\tilde A^{\mu\alpha}\tau_\alpha\tilde A^{\nu\beta}\zeta_\nu\kappa_{\beta\gamma}\dpar_\mu \tilde A^{\kappa\gamma}\zeta_\kappa\Big.\\
 &~~~~~~~~~~~\Big. 
 -\tilde A^{\mu\alpha}\zeta_\mu\tilde A^{\nu\beta}\zeta_\nu \tilde A^{\kappa\gamma}\zeta_\kappa f_{\alpha\beta\gamma}\Big)\\
 &~~+\frac{1}{4!}\Big(
 -\tilde A^{\mu\alpha}\zeta_\mu\kappa_{\alpha\beta}(\dpar_\nu \tilde A^{\kappa\beta})\zeta_\kappa\tilde A^{\lambda\gamma}\zeta_\lambda\kappa_{\gamma\delta}\tilde A^{\nu\delta}
 +2\tilde A^{\mu\alpha}\zeta_\mu\kappa_{\alpha\beta}\tilde A^{\nu\gamma}\zeta_\nu\kappa_{\gamma\delta}(\dpar_\kappa\tilde A^{\lambda\delta})\zeta_\lambda \tilde A^{\kappa\beta}\Big.\\
 &~~~~~~~~~~~\Big. 
 -2\tilde A^{\mu\alpha}\zeta_\mu\kappa_{\alpha\beta} \tilde A^{\nu\gamma}\zeta_\nu\kappa_{\gamma\delta}(\dpar_\kappa\tilde A^{\lambda\beta})\zeta_\lambda\tilde A^{\kappa\delta}
 -\tilde A^{\mu\alpha}\tilde A^{\kappa\beta}\tilde A^{\nu\gamma}\partial_\kappa \tilde A^{\lambda \delta} \kappa_{\alpha\beta}\kappa_{\gamma\delta} \zeta_\mu\zeta_\nu\zeta_\lambda
 \Big)~.
\end{aligned}
\end{equation}
As we can see from the definition of the Poisson structure, the exponential series terminates after the fourth order. Furthermore we observe that we can rewrite the result in a more elegant way by introducing heterotic fluxes of the type $\tilde J_\mu{}^{\nu\alpha}, \tilde K^{\alpha\beta\mu}, \tilde G^{\alpha\mu\nu}$ as well as the so-called nongeometric\footnote{We use the term ``nongeometric'' to make contact with the literature. In our case, the geometric interpretation is the one we describe, i.e.~as twist via a canonical transformation.} fluxes $Q_\mu{}^{\nu\kappa}$ and $R^{\mu\nu\rho}$:
\begin{equation}
\begin{aligned}
  \tilde \CQ =\;& ~\xi^\mu p_\mu - \tfrac{1}{2}\beta^{\mu\nu}\zeta_\mu p_\nu - \tfrac{1}{3!}\,f^{\alpha\beta\gamma}\tau_\alpha\tau_\beta\tau_\gamma - \tilde A^{\mu\alpha}p_\mu \tau_\alpha - \tfrac{1}{2} \tilde A^{\mu\alpha} \tilde A^{\nu\beta}\kappa_{\alpha\beta}\zeta_\mu p_\nu \\
  &+ \tilde J_\mu{}^{\nu\alpha}\,\xi^\mu\zeta_\nu\tau_\alpha + (\tilde K^{\alpha\beta\mu}-\tfrac{1}{2}\tilde A^{\mu\gamma} f^{\alpha\beta\delta}\kappa_{\gamma\delta})\zeta_\mu \tau_\alpha\tau_\beta \\
  &+ (\tilde G^{\alpha \mu\nu}-\tfrac{1}{2} \tilde A^{\mu\beta}\tilde A^{\nu\gamma} \kappa_{\beta\delta}\kappa_{\gamma\epsilon} f^{\alpha\delta\epsilon})\tau_\alpha \zeta_\mu\zeta_\nu \\
  &+ Q_{\mu}{}^{\nu\kappa} \,\xi^\mu\zeta_\nu\zeta_\kappa + (R^{\mu\nu\rho} - \tfrac{1}{3!}\, \tilde A^{\mu\alpha}\tilde A^{\nu\beta}\tilde A^{\rho\gamma} f_{\alpha\beta\gamma})\zeta_\mu\zeta_\nu\zeta_\rho \;.
  \end{aligned}
\end{equation}
We refer the reader to the existing literature, e.g.~\cite{Blumenhagen:2014iua} for a definition (with sign and prefactor conventions slightly different to ours) and interpretation of the heterotic fluxes in terms of DFT and their reduction to heterotic Generalized Geometry. To sum up, we can give an algebraic interpretation of the appearance of fluxes in heterotic generalized geometry: They arise via the action of canonical transformations on the original homological vector field.

\subsection{Generalized Metric}\label{ssec:generalized_metric}

Generalized vectors on $\CW_2(M,P)$ are sections of $E$, which carries an action of the group $\sO(D,D+n)$, where $D$ and $n$ are the dimensions of $M$ and the gauge group $\sG$ with Lie algebra $\frg=\sLie(\sG)$, respectively. There is an $\sO(D,D+n)$ invariant metric, which reads as 
\begin{equation}
 \eta_{MN}=\left(\begin{array}{ccc} 0 & 0 & \unit \\ 0 & \kappa & 0 \\ \unit  & 0 & 0\end{array}\right)
\end{equation}
with $\kappa$ the Killing form on $\frg$. The generalized metric on $\CW_2(M,P)$ is the object that brings together the metric $g_{\mu\nu}$ on $M$, the metric $\kappa_{\alpha\beta}$ on $\frg$ as well as the one-form potential $A$ and the two-form potential $B$ in a covariantly and homogeneously transforming object. Just as the Hamiltonian of the homological vector field, we can construct the generalized metric by acting with canonical transformations on the metric $\CH_0$ with $A=B=0$, which reads as
\begin{equation}
 (\CH_{0,MN})=\begin{pmatrix}
               g_{\mu\nu} & 0 & 0\\
               0 & \kappa_{\alpha\beta} & 0 \\
               0 & 0 & g^{\mu\nu} 
              \end{pmatrix}~.
\end{equation}
We regard this matrix as the tensor $\CH_0=\CH_{0,MN}\xi^M\otimes \xi^N$, cf.~\eqref{eq:xiM}. Using the Poisson bracket extended to tensors according to~\eqref{eq:ext_Poisson}, we find that 
\begin{equation}\label{eq:metric_GG}
 \begin{aligned}
 \CH&=\de^{A^\alpha_\mu \xi^\mu\tau_\alpha}\acton\left(\de^{\tfrac12 B_{\mu\nu}\xi^\mu\xi^\nu}\acton\CH_0\right)\\
 (\CH_{MN})&=\begin{pmatrix}
              g_{\mu\nu}+A_\mu^\alpha\kappa_{\alpha\beta}A_\nu^\beta+\CB_{\kappa\mu}g^{\kappa\lambda}\CB_{\lambda\nu} & A^\alpha_\mu\kappa_{\alpha\beta}-\CB_{\mu\nu}g^{\nu\kappa}A_\kappa^\alpha \kappa_{\alpha\beta} & \CB_{\mu\kappa}g^{\kappa\nu}\\
               \kappa_{\beta\alpha}A^\alpha_\mu - \kappa_{\beta\alpha} A^\alpha_\kappa g^{\kappa\nu}\CB_{\mu\nu} &\kappa_{\alpha\beta}+\kappa_{\alpha\gamma}A^\gamma_\mu g^{\mu\nu} A_\nu^\delta \kappa_{\delta\beta}  & -\kappa_{\alpha\beta}A^\beta_\mu g^{\mu\nu} \\
               \CB_{\nu\kappa} g^{\kappa\mu}  & -g^{\mu\nu}A_\mu^\beta \kappa_{\alpha\beta}  & g^{\mu\nu} 
              \end{pmatrix}~,
 \end{aligned}
\end{equation}
where we defined for convenience $\CB_{\mu\nu} := B_{\mu\nu}-\tfrac{1}{2}\kappa_{\alpha\beta}A^\alpha_\mu A^\beta_\nu$. In order to compute the twist, we use the ordering $A^\alpha_\mu \xi^\mu\tau_\alpha$ as a \emph{definition} of the twist function. This allows us to perform the calculation using solely the tensor product rule~\eqref{eq:ext_Poisson}. The possibility of constructing the generalized metric in this way via symplectomorphisms does not seem to have been noticed before, not even for the exact Courant algebroid.

Using the second type of canonical transformations parameterized by $(\tilde A,\beta)$ and analogously defining the twist function to be $\tilde A^{\mu\alpha}\zeta_\mu \tau_\alpha $ we obtain 
\begin{equation}
 \begin{aligned}
 \tilde \CH&=\de^{\tilde A^{\mu\alpha} \zeta_\mu\tau_\alpha}\acton\left(\de^{\tfrac12 \beta^{\mu\nu}\zeta_\mu\zeta_\nu}\acton\CH_0\right)\\
 (\tilde \CH_{MN})&=\\
 &\begin{pmatrix}
  g_{\mu\nu} & -g_{\mu\nu}\tilde A^{\nu\beta}\kappa_{\beta\alpha} & g_{\mu\kappa}\tilde \CB^{\kappa\nu} \\
 -\kappa_{\alpha\beta}\tilde A^{\nu\beta} g_{\nu\mu} & \kappa_{\alpha\beta} +g_{\rho\nu} \tilde A^{\rho\gamma}\tilde A^{\nu\delta}\kappa_{\alpha\gamma}\kappa_{\delta\beta} & \kappa_{\alpha\beta}\tilde A^{\mu\beta} - \tfrac{1}{2} \kappa_{\alpha\beta}g_{\kappa\nu} \tilde A^{\nu\beta}\tilde \CB^{\mu\kappa} \\
 g_{\nu\kappa}\tilde \CB^{\kappa\mu} & \tilde A^{\mu\beta}\kappa_{\beta\alpha} - \tfrac{1}{2}\kappa_{\alpha\beta}\tilde A^{\nu\beta}g_{\kappa\nu}\tilde \CB^{\mu\kappa} & g^{\mu\nu}+ \kappa_{\alpha\beta} \tilde A^{\mu\alpha}\tilde A^{\nu\beta} + g_{\rho\sigma}\tilde \CB^{\mu\rho}\tilde \CB^{\nu\sigma}
              \end{pmatrix}~,
 \end{aligned}
\end{equation}
where for convenience we defined the tensor $\tilde \CB^{\mu\nu}:=\beta^{\mu\nu}-\tfrac{1}{2} \tilde A^{\mu\alpha}\tilde A^{\nu\beta} \kappa_{\alpha\beta}$. We see again the power of the twist by canonical transformations: The full heterotic generalized metrics (i.e.~including either $(A,B)$ or $(\tilde A,\beta)$) arise from the original $\CH_0$ by applying the exponentiated infinitesimal canonical shifts including the respective parameters.

\subsection{Generalized Riemannian geometry}\label{ssec:GG_Riemannian}

So far, rephrasing the heterotic Courant algebroid as a symplectic N$Q$-manifold gave us a transparent description of the symmetries and the differential geometry of heterotic supergravity. A further advantage of the formulation is that it allows us to lift the expressions found in~\cite{Deser:2016qkw} for torsion and Riemann tensor for type II Double Field Theory to heterotic Generalized Geometry. In the following, we briefly review and specialize the constructions of~\cite{Deser:2016qkw}; for another approach to connections on Courant algebroids, see also~\cite{Garcia-Fernandez:2013gja, Jurco:2015bfs,Jurco:2016emw,Jurco:2017gii}.

A symplectic N$Q$-manifold $\CM$ of degree~$n$ comes with a set of generalized vector fields $\CCX(\CM)$, which are the functions of degree~$n-1$. In this paper, we are exclusively interested in the case $n=2$, and for simplicity, we restrict ourselves to this case in the following. 

We define an {\em extended covariant derivative} on $\CM$ as a Hamiltonian function $\nabla_X$ parameterized by an element $X\in \CCX(\CM)$ such that 
\begin{equation}
\{\nabla_{fX+Y},Z\}=f\{\nabla_X,Z\}+\{\nabla_Y,Z\}\eand\{\nabla_X,fY\}=\{Q X,f\}Y+f\{\nabla_X,Y \}
\end{equation}
holds for all $f\in \CCC^\infty(\CM)$ and $X,Y,Z\in \CCX(\CM)$.

Using this notion of covariant derivative, we can define an extended torsion tensor
\begin{equation}
  \label{eq:extorsion}
  \CT(X,Y,Z):=3\Bigl\{X,\{\nabla_{Y},Z\}\Bigr\}_{[X,Y,Z]}+\tfrac{1}{2}\left(\{X,\{QZ,Y\}\}-\{Z,\{QX,Y\}\}\right)~,  
\end{equation}
where $[X,Y,Z]$ denotes weighted total antisymmetrization in $X,Y,Z$, as well as an extended Riemann tensor 
\begin{equation}
  \label{eq:exCurvature}
  \begin{aligned}
    \CR(X,Y,Z,W):=\; &\tfrac{1}{2}\Bigl(\Bigl\{\bigl\{\{\nabla_X,\nabla_Y\} -\nabla_{\mu_2(X,Y)},Z\bigr\},W\Bigr\} -( Z\leftrightarrow W)   \\
& + \Bigl\{\bigl\{\{\nabla_Z,\nabla_W\}-\nabla_{\{\nabla_Z,W\} - \{\nabla_W,Z\}},X\bigr\},Y\Bigr\} - (X\leftrightarrow Y)\Bigr)\;.
\end{aligned}
\end{equation}
Both $\CT$ and $\CR$ are indeed extended tensor fields and they are $\CC^\infty_0(\CM)$-linear in all their arguments, while the naive definitions of $\CT$ and $\CR$ do not enjoy this property. The torsion~\eqref{eq:extorsion} is the well-known Gualtieri-torsion in case of Generalized Geometry and it is this object which we use to generalize torsion to the case of heterotic Generalized Geometry. In case of vanishing torsion, in~\cite{Hohm:2011ex} an explicit form of the generalized Riemann tensor was determined. Locally, the covariant derivative operator has the form
\begin{equation}
\nabla_X =\; X^Mp_M - \tfrac{1}{2} X^M\Gamma_{MNK} \xi^N\xi^K\;,
\end{equation}
and the scalar curvature corresponding to~\eqref{eq:exCurvature} takes the form 
\begin{equation}
\CR = \; 2 R_0 + \Gamma^{MNK}\Gamma_{MNK}\;,
\end{equation}
where $R_0$ denotes the combination of connection coefficients corresponding to the standard Riemannian scalar curvature, i.e.\footnote{We raise and lower indices by contracting with the constant metric $\eta_{MN}$.} 
\begin{equation}
\label{genscalcurv}
R_0 =\; 2(\partial^{[I}\Gamma^{J]}{}_{IJ} + \Gamma_{[I}{}^{JK}\Gamma_{J]K}{}^I)\;.
\end{equation}
Already in Generalized Geometry, Levi--Civita connections are not unique and there is no analogue of the Koszul formula to determine the connection coefficients. However all possible choices result in the same scalar curvature. An analysis of the constrains on $\Gamma_{MNK}$ in~\cite{Hohm:2011si} provided a suitable choice to express the connection coefficients in terms of the generalized metric ${\CH}_{MN}$ leading to the Lagrangian
\begin{equation}\label{eq:actionHSG_2}
\begin{aligned}
S = \; \int \,dx\, e^{2d}\Bigl(&\tfrac{1}{8}\CH^{IJ}\partial_I \CH^{KL}\partial_J\CH_{KL} - \tfrac{1}{2}\CH^{MI}\partial_I \CH^{KJ}\partial_J\CH_{MK} \\
&-2\partial_I d \,\partial_J \CH^{IJ} + 4\CH^{IJ}\partial_I d\,\partial_J d\Bigr)\;.
\end{aligned}
\end{equation}
For the case of heterotic supergravity, it was shown in~\cite{Hohm:2011ex} that taking $e^{-2d}=\sqrt{g}e^{-2\phi}$ in~\eqref{eq:actionHSG_2} leads to the action~\eqref{hetaction}.

\section{Heterotic Double Field Theory}\label{sec:hDFT}

\subsection{The symplectic pre-N\texorpdfstring{$Q$}{Q}-manifold}

We saw above that the transitive Courant algebroid $\CW_2(M,P)$ captures nicely the symmetries of the non-abelian gerbe arising in the low-energy sector of the heterotic string. In this section, we extend this picture to Double Field Theory, using the formalism developed in~\cite{Deser:2016qkw}. 

We are after an extension of the notion of Courant algebroid describing the local symmetries (as well as their actions) of the bosonic sector of heterotic supergravity in a manifestly T-dual way. The relevant object is therefore not the Courant algebroid itself, but its associated Lie $2$-algebra. We follow~\cite{Deser:2016qkw} and observe that the condition $Q^2=0$ is sufficient but not necessary for the associated Lie $2$-algebra to exist, and we can therefore relax this condition. Preserving the associated Lie $2$-algebra will then require restricting to certain subsets of functions on the extended Courant algebroid. This motivates the following definition of symplectic pre-N$Q$-manifolds.

A {\em symplectic pre-N$Q$-manifold}~\cite{Deser:2016qkw} is a symplectic N-manifold endowed with a vector field $Q$ of degree~1 satisfying $\CL_Q\omega=0$. In particular, a symplectic pre-N$Q$-manifold $(\CM,Q,\omega)$ with $Q^2=0$ is a symplectic N$Q$-manifold. 

From our discussion of the Generalized Geometry underlying heterotic supergravity, together with the symplectic pre-N$Q$-manifold relevant in type II DFT as developed in~\cite{Deser:2016qkw}, it is now reasonably clear what the relevant symplectic pre-N$Q$-manifold is. Just as in type II DFT, we construct it by reduction of a Courant algebroid.

We start from a symplectic Lie 2-algebroid with underlying graded manifold 
\begin{equation}
 T^*[2]T[1](T^*M\times \frg)~~\cong~~ T^*[2]T[1]T^*M~\times~T^*[2]T[1]\frg~,
\end{equation}
where the first factor is indeed the Vinogradov algebroid\footnote{see appendix} $\CV_2(T^*M)=T^*[2]T[1](T^*M)$ and the second factor $T^*[2]T[1]\frg$ has the same underlying graded manifold as $\CV_2(\frg)$, but different homological vector field. Explicitly, we introduce coordinates $(x^m,\xi^m,\zeta_m,p_m)$, $m=1,\dots,2D$ of degrees $(0,1,1,2)$ on $\CV_2(T^*M)$ and $(y_\alpha,\psi_\alpha,\phi^\alpha,q^\alpha)$ of degrees $(0,1,1,2)$ on $T^*[2]T[1]\frg$. Since we anticipate that this structure should reduce to the transitive Courant algebroid describing twisted string structures, we put $\frg=\ao(D,D)\times \frh\cong\apin(D,D)\times \frh$, where $\frh$ should be thought of as either $\aso(32)$ or $\ae_8\times \ae_8$. The canonical symplectic structure on $T^*[2]T[1](T^*M\times \frg)$ reads as
\begin{equation}
 \omega=\dd x^m\wedge \dd p_m+\dd \xi^m\wedge \dd \zeta_m+\dd y_\alpha\wedge \dd q^\alpha+\dd \psi_\alpha\wedge \dd\phi^\alpha
\end{equation}
and the homological vector field is has Hamiltonian
\begin{equation}
 \CQ=\sqrt{2}(\xi^m p_m+\psi_\alpha q^\alpha)-\frac{(\sqrt{2})^3}{3!}f^{\alpha\beta\gamma}\psi_\alpha\psi_\beta\psi_\gamma~.
\end{equation}
Note that this Hamiltonian differs in the last summand from that on $\CV_2(T^*M\times \frg)$.

To reduce the pre-N$Q$-manifold relevant for heterotic DFT, we introduce the new coordinates of degree~1,
\begin{equation}
\begin{aligned}
 \theta^m&=\frac{1}{\sqrt{2}}(\xi^m+\eta^{mn}\zeta_n)~,~~~&\beta^m&=\frac{1}{\sqrt{2}}(\xi^m-\eta^{mn}\zeta_n)~,\\
 \tau_\alpha&=\frac{1}{\sqrt{2}}(\psi_\alpha+\kappa_{\alpha\beta}\phi^\beta)~,~~~&\sigma_\alpha&=\frac{1}{\sqrt{2}}(\psi_\alpha-\kappa_{\alpha\beta}\phi^\beta)~,
\end{aligned}
\end{equation}
where $\kappa_{\alpha\beta}$ is the Killing metric on $\frg$. The reduced pre-N$Q$-manifold $\CF_2(M,\frg)$ is now obtained by putting $\beta^m=0$ and $\sigma_\alpha=0$.

Explicitly, $\CF_2(M,\frg)$ is described by coordinates $(x^m,y_\alpha, \theta^m,\tau_\alpha,p_m,q^\alpha)$ of degrees $(0,0,$ $1,1,2,2)$ and carries the symplectic form
\begin{equation}
\omega=\dd x^m\wedge \dd p_m+\tfrac12 \eta_{mn}\dd \theta^m\wedge \dd \theta^n+\dd y_\alpha\wedge \dd q^\alpha+\tfrac12 \kappa^{\alpha\beta}\dd \tau_\alpha\wedge \dd \tau_\beta~,
\end{equation}
where the matrix $\kappa^{\alpha\beta}$ is the inverse of $\kappa_{\alpha\beta}$. The corresponding Poisson bracket reads as 
\begin{equation}
\begin{aligned}
 \{f,g\}:=&f\overleftarrow{\der{p_m}}\overrightarrow{\der{x^m}} g-f\overleftarrow{\der{x^m}}\overrightarrow{\der{p_m}} g-f\overleftarrow{\der{\theta^m}}\eta^{mn}\overrightarrow{\der{\theta^n}} g\\
 &+f\overleftarrow{\der{q^\alpha}}\overrightarrow{\der{y_\alpha}} g-f\overleftarrow{\der{y_\alpha}}\overrightarrow{\der{q^\alpha}} g-f\overleftarrow{\der{\tau_\alpha}}\kappa_{\alpha\beta}\overrightarrow{\der{\tau_\beta}} g
\end{aligned}
\end{equation}\label{eq:E2_Poisson}
for $f,g\in\CC^\infty(\CF_2(M,\frg))$. The Hamiltonian $\CQ$ and its vector field $Q$ of $\CV_2(T^*M\times \frg)$ restrict to 
\begin{equation}
 \CQ=\theta^m p_m+\tau_\alpha q^\alpha-\tfrac{1}{3!}f^{\alpha\beta\gamma}\tau_\alpha\tau_\beta\tau_\gamma
\end{equation}
and
\begin{equation}
 Q=\theta^m\der{x^m}+p_m\eta^{mn}\der{\theta^n}+\tau_\alpha\der{y_\alpha}+q^\alpha\kappa_{\alpha\beta}\der{\tau_\beta}+\frac{1}{2} \tau_\alpha\tau_\beta f^{\alpha\beta\gamma}\kappa_{\gamma\delta}\der{\tau_\delta}
\end{equation}
with
\begin{equation}
 Q^2=p_m \eta^{mn}\der{x^n}+q^\alpha\kappa_{\alpha\beta}\der{y_\beta}+\tfrac12 \tau_\alpha\tau_\beta f^{\alpha\beta\gamma}\kappa_{\gamma\delta}\der{y_\delta}+q^\alpha\kappa_{\alpha\beta}f^{\beta\gamma\delta}\tau_\gamma\kappa_{\delta\eps}\der{\tau_\eps}\neq 0~.
\end{equation}
This symplectic pre-N$Q$-manifold $\CF_2(M,\frg)$ is now suitable for a description of the symmetries of heterotic DFT, as we shall see in the following. We already note that $\CF_2(M,\frg)$ has the right dimension in degree~0 and as expected, putting $\frg=*=\{0\}$ yields the pre-N$Q$-manifold $\CE_2(M)$ used in~\cite{Deser:2016qkw} to capture the symmetries of type II DFT.

To render the whole construction $\sO(D,D+n)$-covariant, one can split $x^m=(x^\mu,x_\mu)$ and introduce combined coordinates $x^M=(x^\mu,y_\alpha,x_\mu)$, the combined metric
\begin{equation}
  (\eta_{MN})=\begin{pmatrix}
         0 & 0 & \unit \\ 0 & \kappa & 0 \\ \unit & 0 & 0
        \end{pmatrix}
\end{equation}
and the combined structure constants
\begin{equation}
 (f_{MNP})=\left\{\begin{array}{l l}
                   f_{\alpha\beta\gamma} & \mbox{~for~}M,N,P=\alpha,\beta,\gamma\\
                   0 & \mbox{~else}~.
                  \end{array}\right.
\end{equation}
Since this might make our constructions slightly more opaque, we refrain from using this formulation.

\subsection{Heterotic C- and D-brackets}

The C- and D-brackets of heterotic Double Field Theory are simply obtained as derived brackets on the symplectic pre-N$Q$-manifold $\CF_2(M,\frg)$, as we shall explain in the following. Extended vector fields are functions on $\CF_2(M,\frg)$ of degree~1 and we have
\begin{equation}
 \CC^\infty_1(\CF_2(M,\frg))=\{X_m\theta^m+X^\alpha\tau_\alpha~|~X_m,X^\alpha\in \CC^\infty(M)\}~.
\end{equation}
Here $m=1,\dots,2\dim(M)$ and $\alpha=1,\dots,\dim(\frg)$ as in the previous section. The generalized Lie derivative, or {\em D-bracket} is now defined by the same algebraic formula as in Generalized Geometry, cf.~section~\ref{ssec:het_Courant_algebroids}:
\begin{equation}\label{eq:pre_D_action}
 \hat \CL_X Y:=[X,Y]_D:=\nu_2(X,Y):=\{QX,Y\}~,~~~X,Y\in \CC^\infty_1(\CF_2(M,\frg))~.
\end{equation}
In components, we have the following formula:
\begin{equation}
\begin{aligned}
 \hat \CL_X Y&=X^m\der{x^m}Y-Y^m\der{x^m}X+X^\alpha\der{y^\alpha}Y-Y^\alpha\der{y^\alpha}X+\theta^m Y^n\der{x^m}X_n+\\
 &~~~+\theta^mY^\alpha\der{x^m}X_\alpha+\tau^\alpha Y^n\der{y^\alpha}X_n+\tau^\alpha Y^\beta\der{y^\alpha}X_\beta+\tau^\alpha f_{\alpha\beta\gamma}X^\gamma Y^\beta~,
\end{aligned}
\end{equation}
where indices are raised and lowered with the $\sO(D,D)$-metric $\eta_{mn}$ or the Killing metric $\kappa_{\alpha\beta}$. This is indeed the generalized Lie derivative of heterotic DFT~\cite{Hohm:2011ex}.

The action~\eqref{eq:pre_D_action} readily generalizes to extended tensor fields $\sT(\CF_2(M,\frg))$, which are a subset of the free tensor algebra $\CT(\CM)$ of $\CC^\infty(\CF_2(M,\frg))$ if we extend the Poisson bracket as follows:
\begin{equation}\label{eq:ext_Poisson2}
 \{f,g\otimes h\}:=\{f,g\}\otimes h+(-1)^{(n-|f|)|g|}g\otimes \{f,h\}~,
\end{equation}
cf.~\cite{Deser:2016qkw}. 

The C-bracket is then simply the antisymmetrization of the D-bracket:
\begin{equation}
 [X,Y]_C:=\mu_2(X,Y):=\tfrac12 (\{QX,Y\}-\{QY,X\})~,~~~X,Y\in \CC^\infty_1(\CF_2(M,\frg))~.
\end{equation}
As easily verified from the algebraic relations for the graded Poisson structure $\{-,-\}$ and $Q$, we have the equivalent relations
\begin{equation}
\begin{aligned}
 \hat \CL_X\hat \CL_Y Z-\hat \CL_Y\hat \CL_X Z&=\hat \CL_{[X,Y]_C} Z~,\\
 \nu_2(X,\nu_2(Y,Z))-\nu_2(Y,\nu_2(X,Z))&=\nu_2(\mu_2(X,Y),Z)
\end{aligned}
\end{equation}
for all $X,Y,Z\in\CC^\infty_1(\CF_2(M,\frg))$.

It is important to note that the C- and D-brackets are just a part of a larger algebraic structure. The C-bracket is a binary bracket belonging to a Lie 2-algebra, and this Lie 2-algebra can be reduced to that of heterotic Generalized Geometry. The D-bracket, on the other hand, describes part of an action of the Lie 2-algebra on the vector space of extended tensor fields. The detailed analysis of these algebraic structures yields the appropriate section condition of Double Field Theory, as we shall discuss next.

\subsection{\texorpdfstring{$L_\infty$}{Linfinity}-structures and the strong section condition}

First, recall from~\cite{Deser:2016qkw} that an {\em $L_\infty$-algebra structure} $\sL(\CM)=\oplus_{i=0}^{n-1} \sL_i(\CM)$ on a pre-N$Q$-manifold $(\CM,\omega,Q)$ of degree~$n$ is a subset $\sL(\CM)\subset \CC^\infty(\CM)$ such that the higher derived brackets~\eqref{eq:ass_L_infty_structure} form an $L_\infty$-algebra. We also demand an action of this $L_\infty$-algebra on extended vector fields $\sL_0(\CM)\subset \CC^\infty_{n-1}(\CM)$. That is, the Dorfman bracket~\eqref{eq:def:Dorfman_bracket} should provide an action of $L_\infty$-algebras of $\sL(\CM)$ on $\sL_0(\CM)$, cf.\ again~\cite{Deser:2016qkw}. Both conditions yield a weakened form of the strong section condition.

The first condition was abstractly analyzed for pre-N$Q$-manifolds of degree~$2$ in~\cite[Theorem~4.7]{Deser:2016qkw}. Let $\sL(\CM)$ be a subset of $\CC^\infty(\CM)$ concentrated in $\sL$-degrees $0$ and $1$ and write $\sL(\CM)=\sL_1(\CM)\oplus \sL_0(\CM)$. Then $\sL(\CM)$ is an $L_\infty$-algebra if and only if
 \begin{equation}\label{eq:conditions_thm_Lie2_subset}
  \begin{aligned}
    \{Q^2f,g\}+\{Q^2g,f\}&=0~,\\
    \{Q^2X,f\}+\{Q^2f,X\}&=0~,\\
    \{\{Q^2X,Y\},Z\}_{[X,Y,Z]}&=0
  \end{aligned}
 \end{equation}
for all $f,g\in \sL_1(\CM)$ and $X,Y,Z\in \sL_0(\CM)$. In our case with $\CM=\CF_2(M,\frg)$, where
\begin{equation}
 \sL_1(\CM)\subset\CC^\infty(M)\eand \sL_0(\CM)\subset\CC^\infty_1(\CM)=\{X_m\theta^m+X^\alpha\tau_\alpha~|~X_m,X^\alpha\in \CC^\infty(M)\}~,
\end{equation}
this translates to the equations
\begin{equation}\label{eq:conditions_thm_Lie2_subset_exp}
\begin{aligned}
  2(\dpar_Mf)(\dpar^Mg)&=0~,\\
  2\Big((\dpar_MX)(\dpar^M f)+X_\beta\fd^{\beta}f\Big)&=0~,\\
  \big((\dpar_MX)(\dpar^MY^L)Z_L+(\fd_\alpha X^L)Y_LZ^\alpha-2(\fd_{\alpha\beta}X)Y^\alpha Z^\beta\big)_{[X,Y,Z]}&=0~,
\end{aligned}
\end{equation}
where $\dpar_M:=(\der{x^m},\der{y_\alpha})$ and $X_M=(X_m,X^\alpha)$, $X^M=(\eta^{mn}X_n,\kappa_{\alpha\beta}X^\beta)$, etc. To improve legibility of the equations, we have used the shorthand notation
\begin{equation}
 \fd^{\alpha\beta} F:=f^{\gamma\alpha\beta}\kappa_{\gamma\delta}\der{y_\delta}F~,~~~\fd^{\alpha} F:=\tau_\beta \fd^{\beta\alpha}F\eand \fd F:=\tau_\alpha \fd^{\alpha}F
\end{equation}
for $F\in \CC^\infty(\CM)$ and allow for raising and lowering of indices with the Killing metric $\kappa_{\alpha\beta}$. The necessary condition that the Dorfman bracket induces an action of $L_\infty$-algebras on $\sL_0$ is gleaned from~\cite[Theorem~4.11]{Deser:2016qkw}. For all $V,W,X,Y\in \sL_0$, we need
\begin{equation}\label{eq:abstract_action_condition}
  \big\{\{Q^2V,W\}-\{Q^2W,V\},X\big\}=0\eand \Big\{\big\{\{Q^2V,W\}-\{Q^2W,V\},QX\big\},Y\Big\}=0~.
\end{equation}
We readily compute 
\begin{equation}\label{eq:temp_result}
\begin{aligned}
 \{Q^2V,W\}-\{Q^2W,V\}&=(\dpar^M V)(\dpar_M W)+(\dpar_M V^L)p^M W_L+\\
 &~~~~~~~~(\fd_\alpha W)V^\alpha+(\fd W_M)V^M+f_{\alpha\beta\gamma}q^\alpha W^\beta V^\gamma-V\leftrightarrow W~,
\end{aligned}
\end{equation}
where $p_M:=(p_m,q_\alpha)$. Using this relation, we find that the first equation of~\eqref{eq:abstract_action_condition} reads as
\begin{equation}\label{eq:Lie_action_on_itself}
\begin{aligned}
2\Big((\dpar_MV)(\dpar^MW^L)X_L+(\fd_\alpha W^L)X_LV^\alpha+(\fd_\alpha V^L)W_LX^\alpha+(\fd_{\alpha\beta}W)V^\alpha X^\beta\Big)&\\
+(\dpar_MX)(\dpar^MV^L)W_L-(\fd_{\alpha\beta}X)V^\alpha W^\beta-V&\leftrightarrow W=0~,
\end{aligned}
\end{equation}
while the second one translates to a more complicated and not very enlightening expression. Recall that the form of the strong section condition usually employed in heterotic DFT as found in~\cite{Hohm:2011ex} is
\begin{equation}\label{eq:standard_section_condition}
 (\dpar_M F)(\dpar^M G)=0\eand \fd_{\alpha\beta} F=0
\end{equation}
for all fields $F$ and $G$. We note that any Poisson bracket of the form 
\begin{equation}
 \{\{Q^2V,W\}-\{Q^2W,V\},F\}
\end{equation}
with $F$ a field vanishes due to~\eqref{eq:temp_result} after imposing condition~\eqref{eq:standard_section_condition} and therefore conditions~\eqref{eq:abstract_action_condition} are automatically satisfied. We stress, however, that~\eqref{eq:standard_section_condition} is not necessary for~\eqref{eq:abstract_action_condition} to be satisfied.

The extended tensors fields $\sT(\CM)$ are a subset of the free tensor algebra $\CT(\CM)$ of $\CC^\infty(\CF_2(M,\frg))$ over $\CC^\infty(T^*M)$, and elements $t\in\sT(\CM)$ satisfy the conditions
\begin{equation}\label{eq:abstract_action_condition_tensors}
 \big\{\{Q^2V,W\}-\{Q^2W,V\},t\big\}=0\eand \Big\{\big\{\{Q^2V,W\}-\{Q^2W,V\},QX\big\},t\Big\}=0~.
\end{equation}
We refrain from giving explicit expressions for these conditions, which, however, can be readily computed using~\eqref{eq:temp_result}. Again, the usual strong section condition of heterotic DFT~\eqref{eq:standard_section_condition} is sufficient for~\eqref{eq:abstract_action_condition_tensors} to hold, but not necessary. Altogether, equations~\eqref{eq:conditions_thm_Lie2_subset_exp},~\eqref{eq:abstract_action_condition} and~\eqref{eq:abstract_action_condition_tensors} form the appropriate and indeed weakened form of the usual strong section condition for heterotic Double Field Theory.

Note that our whole discussion reduces to that of type II  Double Field Theory upon putting $\frg=*=\{0\}$. In this context, another condition was suggested in the literature~\cite{Geissbuhler:2013uka}, known as the {\em closure constraint}. This constraint amounts to the fact that the commutator of two generalized Lie derivatives closes to a generalized Lie derivative. In our framework, this amounts to one of the conditions that the generalized Lie derivative forms a representation of the underlying Lie 2-algebra of symmetries on the generalized vector fields, which is encoded in equation~\eqref{eq:Lie_action_on_itself}. (We only consider generalized tensors, while~\cite{Geissbuhler:2013uka} discusses also tensor densities.) Let us stress, however, that for a full description one needs an underlying Lie 2-algebra (our conditions~\eqref{eq:conditions_thm_Lie2_subset}) as well as the fact that the generalized Lie derivative forms a representation of this Lie 2-algebra (our conditions~\eqref{eq:abstract_action_condition}).

\subsection{Examples and reduction to Generalized Geometry}

The canonical example of an $L_\infty$-structure on $\CF_2(M,\frg)$ is certainly the subset $\sL$ of functions in $\CC^\infty(\CF_2(M,\frg))$, which are constant in $x_\mu$, $p^\mu$, $y_\alpha$ and $q^\alpha$. For such functions $F,G$, we have
\begin{equation}
\dpar_M F~\dpar^M G=0~,
\end{equation}
and our section conditions~\eqref{eq:conditions_thm_Lie2_subset} and~\eqref{eq:abstract_action_condition} are automatically satisfied. Note also that tensors, i.e.\ elements of the free tensor algebra $\CT(\CF_2(M,\frg))$, which are constant in the above variables automatically satisfy~\eqref{eq:abstract_action_condition}.

This is not surprising, because by choosing this $L_\infty$-structure and the corresponding tensors, we effectively reduced the symplectic pre-N$Q$-manifold $\CF_2(M,\frg)$ to the heterotic Courant algebroid $\CW_2(M)$ with $F=H=0$, which does not require any further conditions. 

\subsection{Twisting}

The generalization of $\CF_2(M,\frg)$ to some symplectic pre-N$Q$-manifold which reduces to the heterotic Courant algebroid $\CW_2(M)$ with non-trivial $F$ and $H$ is now the next task. Clearly, one needs to modify or twist the vector field $Q$ by adding terms depending on topological data, which reduces to the topological data $F$ and $H$ in Generalized Geometry. This issue is also closely linked to the problem of finding the global picture in heterotic Double Field Theory. Once an appropriate twist is found, we can patch local descriptions together, avoiding the problems pointed out in~\cite{Papadopoulos:2014mxa}.

Note that we would like the $L_\infty$-structure $\sL$ to remain the same before and after twisting. We simply follow~\cite{Deser:2016qkw} and define a twist of $\CF_2(M,\frg)$ that respects the $L_\infty$-structure $\sL$ as the vector field $Q_T$ with Hamiltonian
\begin{equation}
 \CQ_T:=\CQ+T~,
\end{equation}
where $T$ is a function on $\CF_2(M,\frg)$ of degree~3 and $Q_T$ satisfies the appropriate section conditions~\eqref{eq:conditions_thm_Lie2_subset} and~\eqref{eq:abstract_action_condition_tensors} on the $L_\infty$-structure $\sL$. 

As an example, consider the $L_\infty$-structure $\sL$ of the previous section, which reduces $\CF_2(M,\frg)$ to $\CW_2(M)$. Then the function
\begin{equation}\label{eq:twist_DFT}
 T=\tfrac12 A^\alpha_\mu\xi^\mu \kappa_{\alpha\beta}f^{\beta\gamma\delta}\tau_\gamma\tau_\delta-\tfrac{1}{2!}F^\alpha_{\mu\nu}\xi^\mu\xi^\nu\tau_\alpha-\tfrac{1}{3!}H_{\mu\nu\kappa}\xi^\mu\xi^\nu\xi^\kappa
\end{equation}
leads to a twist of $\CF_2(M,\frg)$ for
\begin{equation}
 \dd A+\tfrac12 [A,A]=F~,~~~\nabla F=0\eand \dd H=(F,F)~,
\end{equation}
cf.~\eqref{eq:GG_twist_condition}. We recover the general heterotic Courant algebroid $\CW_2(M)$ after restricting the functions on $\CF_2(M,\frg)$ to $\sL$. 

Note that the twist~\eqref{eq:twist_DFT} is the same as the one we used in Generalized Geometry~\eqref{eq:twist_gg}. This twist continues to work, since the actions
\begin{equation}
 \de^{A^\alpha_\mu \xi^\mu\tau_\alpha}\acton \eand \de^{\tfrac12 B_{\mu\nu}\xi^\mu\xi^\nu}\acton
\end{equation}
defined via the extended Poisson bracket~\eqref{eq:ext_Poisson2} are essentially the same in Generalized Geometry and DFT. Therefore, we can again derive the generalized metric analogously to section~\ref{ssec:generalized_metric} and the result will be the same.

One might be tempted to consider the action of all types of canonical transformations $(A,B,\tilde A,\beta)$ to obtain expressions for the Double Field Theory fluxes and their Bianchi identities. While this works in the case $A=\tilde A=0$ (and the results are given in~\cite{Heller:2016abk}), the computation does not seem to extend to heterotic Double Field Theory in a straightforward manner. This is due to the non-trivial mixing of the various twist components, which turn the exponentials into infinite power series.

\subsection{Torsion and Riemann tensor}

Recall that for heterotic Generalized Geometry, encoding the symmetry structure in a symplectic N$Q$-manifold of degree~2 allowed us to lift results on the torsion and Riemann tensors from ordinary Generalized Geometry to the heterotic case. The same holds true for Double Field Theory, and we can use the structures on $\CF_2(M,\frg)$ to write down torsion and Riemann tensors, which are algebraically identical to the ones of ordinary DFT, ordinary Generalized Geometry and heterotic Generalized Geometry.

Given extended vector fields $X,Y,Z,W\in \CC^\infty_1(\CF_2(M,\frg))$, extended torsion and Riemann tensors can be defined as
\begin{equation}
\begin{gathered}
  \CT(X,Y,Z):=3\Bigl\{X,\{\nabla_{Y},Z\}\Bigr\}_{[X,Y,Z]}+\tfrac{1}{2}\left(\{X,\{QZ,Y\}\}-\{Z,\{QX,Y\}\}\right)~,\\
  \begin{aligned}
    \CR(X,Y,Z,W):=\; &\tfrac{1}{2}\Bigl(\Bigl\{\bigl\{\{\nabla_X,\nabla_Y\} -\nabla_{\mu_2(X,Y)},Z\bigr\},W\Bigr\} -( Z\leftrightarrow W)   \\
& + \Bigl\{\bigl\{\{\nabla_Z,\nabla_W\}-\nabla_{\{\nabla_Z,W\} - \{\nabla_W,Z\}},X\bigr\},Y\Bigr\} - (X\leftrightarrow Y)\Bigr)\;.
\end{aligned}
\end{gathered}
\end{equation}
These expressions are indeed tensors and in particular, they are $\CC^\infty(M)$-linear in each slot, if we use the first two constraints of~\eqref{eq:conditions_thm_Lie2_subset} on the algebra of functions on the base manifold. We emphasize the advantage of reformulating heterotic Double Field Theory in our language: The expressions for torsion and curvature have the same form as in section~\ref{ssec:GG_Riemannian} for heterotic Generalized Geometry. It is the choice of the underlying pre-N$Q$-manifold where the difference of Generalized Geometry and Double Field Theory is rooted. Proceeding in the same way as in section~\ref{ssec:GG_Riemannian}, we arrive at the following Gualtieri-torsion:
\begin{equation}
\CT(X,Y,Z) =\, X^M Y^N Z^K(\Gamma_{MNK} - \Gamma_{NMK} + \Gamma_{KMN})\;,
\end{equation}
which we set to zero in order to get an explicit form for the curvature operator. Contracting the latter in the way to get the scalar curvature (using the flat $\eta_{MN}$) we arrive at
\begin{equation}
\CR = \; 2 R_0 + \Gamma^{MNK}\Gamma_{MNK}\;,
\end{equation}
where again $R_0$ denotes the standard combination of connection coefficients~\eqref{genscalcurv} leading to ordinary scalar curvature. Proceeding now as was done already in the Double Field Theory literature, i.e.~taking the explicit form of $\Gamma_{MNK}$ determined in~\cite{Hohm:2011si}, the heterotic Double Field Theory action is obtained, which has the same form as in Double Field Theory but where the extended metric~\eqref{eq:metric_GG} is used (depending on the extended set of spacetime coordinates).

\section{\texorpdfstring{$\alpha'$}{alpha'}-corrections}\label{sec:alphaprime}

There is an alternative formulation of the Generalized Geometry and Double Field Theory underlying heterotic string theory in which the set of generalized vectors is the same as in the bosonic case (or the bosonic part of type II string theory). In these descriptions, the pairing, the generalized Lie derivative and the Courant bracket (or C-bracket in the DFT case) receive corrections to linear order in $\alpha'$. In this last section, we briefly recall the interpretation of~\cite{Coimbra:2014qaa} of these corrections for Generalized Geometry, show that this fits our formalism and extend the discussion to Double Field Theory.

\subsection{Generalized Geometry}

Consider two generalized vector fields $X,Y$, which are sections of the ordinary generalized tangent bundle $T\FR^d\oplus T^*\FR^d$. For simplicity, we discuss the untwisted case with $H=0$. The pairing, generalized Lie bracket and Courant bracket receive $\alpha'$-corrections, which read as
\begin{equation}\label{eq:corrections}
 \begin{aligned}
  \langle X,Y\rangle_{\alpha'}&=\langle X,Y \rangle_0-\alpha'\dpar_\mu X^\nu\dpar_\nu Y^\mu~,\\
  \nu_2(X,Y)_{\alpha'}&=\nu_2(X,Y)_0-\alpha' \dpar_\mu Y^\nu \dd \dpar_\nu X^\mu~,\\
  \mu_2(X,Y)_{\alpha'}&=\mu_2(X,Y)_0-\tfrac12 \alpha'(\dpar_\mu Y^\nu\dd\dpar_\nu X^\mu-\dpar_\mu X^\nu\dd\dpar_\nu Y^\mu)~,
 \end{aligned}
\end{equation}
cf.~\cite{Coimbra:2014qaa,Hohm:2013jaa}. We now switch to the N$Q$-manifold picture, in which the pairing originates from the Poisson structure on a symplectic N$Q$-manifold and the generalized Lie derivative is a derived bracket. The relevant N$Q$-manifold is $\CM=T^*[2]T[1]M$ with coordinates $(x^\mu,\xi^\mu,\zeta_\mu,p_\mu)$ of degrees~$0,1,1,2$, respectively and the homological vector field and the symplectic form read as
\begin{equation}
 Q=\xi^\mu\der{x^\mu}+p_\mu\der{\zeta_\mu}\eand \omega=\dd x^\mu\wedge \dd p_\mu+\dd \xi^\mu\wedge \dd \zeta_\mu~.
\end{equation}
The uncorrected Lie derivative and Courant brackets are then given by
\begin{equation}
 \nu_2(X,Y)_0=\{QX,Y\}\eand \mu_2(X,Y)_0=\tfrac12\big(\{QX,Y\}-\{QY,X\}\big)~,
\end{equation}
cf.~\eqref{eq:Dorfman_bracket}~and~\eqref{eq:L_infty_brackets} and~\eqref{eq:Dorfman_bracket_gen}. In this framework, all the $\alpha'$-corrections can be absorbed in the following deformation of the Poisson bracket within the space of bi-differential operators:
\begin{equation}\label{eq:def_Poisson_GG}
 \{f,g\}_{\alpha'}=\{f,g\}+\alpha'\left(f \overleftarrow{\der{x^\mu}}\overleftarrow{\der{\zeta_\nu}}\overrightarrow{\der{x^\nu}}\overrightarrow{\der{\zeta_\mu}}g\right)~.
\end{equation}
In particular, we have
\begin{equation}
 \begin{aligned}
  \langle X,Y\rangle_{\alpha'}&=\{X,Y\}_{\alpha'}~,\\
  \nu_2(X,Y)_{\alpha'}&=\{QX,Y\}_{\alpha'}~,\\
  \mu_2(X,Y)_{\alpha'}&=\tfrac12\big(\{QX,Y\}_{\alpha'}-\{QY,X\}_{\alpha'}\big)~.
 \end{aligned}
\end{equation}
This observation is due to~\cite{Deser:2014wva}, where this deformation of the Poisson bracket is interpreted as a star commutator with an underlying star product originating from a Poisson tensor of inhomogeneous degree.

The $\alpha'$ deformations however find another, possibly more natural interpretation as an indicator that the generalized tangent bundle should be extended by the frame bundle~\cite{Coimbra:2014qaa}. Recall that a diffeomorphism induces a local transformation of the frame bundle: given a set of orthogonal basis vectors $\der{x^\mu}$, they transform under an infinitesimal diffeomorphism parameterized by a generalized vector $X$ according to $\CL_X \der{x^\mu}$, which corresponds to a matrix element $\langle \dd x^\nu,\CL_X \der{x^\mu}\rangle=-\dpar_\mu X^\nu$. This matrix element should be regarded as a section of the subbundle $\ad\,\ao(d)$ of $\ad\,\frg$ in the heterotic Courant algebroid $E$ defined in section~\ref{ssec:het_Courant_algebroids}. The pairing on $E$, where the Killing form $\kappa$ is simply $-\alpha'\tr(-)$, then yields the $\alpha'$-corrections in~\eqref{eq:corrections}:
\begin{equation}
\begin{aligned}
 \langle X,Y \rangle_{\alpha'}&=\langle X+(\dpar_\mu X^\nu),Y+(\dpar_\mu Y^\nu)\rangle_E\\
 &=\langle X,Y\rangle_0+\kappa((\dpar_\mu X^\nu),(\dpar_\mu Y^\nu))=\langle X,Y\rangle_0-\alpha'\dpar_\mu X^\nu \dpar_\nu Y^\mu~.
\end{aligned}
\end{equation}
This accounts for all $\alpha'$-corrections in~\eqref{eq:corrections}.

Altogether, we come to the same conclusion as~\cite{Coimbra:2014qaa}: the appearance of $\alpha'$-corrections in the underlying algebraic structures suggests an extension of the generalized tangent bundle. We do not believe that an analysis based on the deformed Poisson bracket~\eqref{eq:def_Poisson_GG} would be successful. In particular, one cannot simply compute $\alpha'$-corrections to the generalized torsion and Riemann tensors introduced in section~\ref{ssec:GG_Riemannian} by replacing the Poisson structure by the $\alpha'$-corrected one, since the latter violates the relevant Leibniz rule and Jacobi identity.

\subsection{Double Field Theory}

Inspired by the results of the previous section, we now turn to the case of heterotic Double Field Theory, where higher derivative corrections to the bilinear pairing and C-bracket were found in~\cite{Hohm:2014eba}. We refer the reader to~\cite{Hohm:2013jaa, Hohm:2013bwa, Hohm:2014xsa} for a detailed study of $\alpha'$ corrections in DFT. Focusing on heterotic DFT, locally they have the form 
\begin{equation}\label{eq:correctionsDFT}
 \begin{aligned}
  \langle X,Y\rangle_{\alpha'}&=\langle X,Y \rangle_0-\alpha'\dpar_m X^n\,\dpar_n Y^m~,\\
  \nu_2(X,Y)_{\alpha'}&=\nu_2(X,Y)_0-\alpha' \dpar_m\dpar_k X^q\,\dpar_q Y^k\,\theta^m~,\\
  \mu_2(X,Y)_{\alpha'}&=\mu_2(X,Y)_0-\tfrac12 \alpha'(\dpar_m\dpar_q X^k \dpar_k Y^q - X\leftrightarrow Y)\theta^m~.
 \end{aligned}
\end{equation}
Again, these are higher derivative corrections and thus an interpretation in terms of a deformation of the original Poisson structure leading to the derived brackets for ordinary DFT suggests itself. Therefore we start with the pre-N$Q$-manifold for DFT introduced in~\cite{Deser:2016qkw}. This is a reduced version of $\CV_2(T^*M)$ with canonical Poisson structure and Hamiltonian function given by
\begin{equation}
\omega =\, dx^m \wedge dp_m +\tfrac{1}{2}\eta_{mn}\,d\theta^m\wedge d\theta^n\;, \qquad \CQ =\,\theta^m p_m\;.
\end{equation}
Motivated by the previous section, it is now easy to see that the appropriate deformation of the resulting Poisson bracket to receive the corrections~\eqref{eq:correctionsDFT} using a derived bracket is
\begin{equation}\label{eq:def_Poisson_DFT}
 \{f,g\}_{\alpha'}=\{f,g\}+\alpha'\left(f \overleftarrow{\der{x^m}}\overleftarrow{\der{\theta_n}}\overrightarrow{\der{x^n}}\overrightarrow{\der{\theta_m}}g\right)~.
\end{equation}
Applying the deformation to the definition of the bilinear pairing immediately gives the desired result, i.e.\ for $X= X^m\theta_m$ and $Y = Y^m\theta_m$ we get
\begin{equation}
\{X,Y\}_{\alpha'} =\, \langle X,Y\rangle_{\alpha'}\;.
\end{equation}
Deforming the Dorfman bracket $\nu_2$ yields
\begin{equation}
\lbrace \lbrace \CQ, X\rbrace_{\alpha'},Y\rbrace_{\alpha'} =\, \nu_2(X,Y) +\alpha'(\dpar_m \dpar^n X^k\,\dpar_n Y^m  - \partial_m \partial^k X^n \partial_n Y^m)\theta_k\;.
\end{equation}
The second term of order $\alpha'$ gives the desired correction term. The first term is not seen by standard Double Field Theory as the contraction of derivatives vanishes due to the strong section condition. Thus, the deformation~\eqref{eq:def_Poisson_DFT} reproduces the Dorfman derivative of Double Field Theory up to the strong section condition. Turning finally to $\mu_2$, we get
\begin{equation}
  \label{Cbracketdeformation}
  \begin{aligned}
    \mu_2(X,Y) =\, \mu_2(X,Y)_0 &+ \alpha'(\dpar_m \dpar^n X^k\, \dpar_n Y^m -\dpar_m \dpar^k X^n\, \dpar_n Y^m)\theta_k \\
    &- \alpha'(\dpar_m \dpar^n Y^k \,\dpar_n X^m - \dpar_m \dpar^k Y^n \,\dpar_n X^m)\theta_k\;.
  \end{aligned}
\end{equation}
Again, as for $\nu_2$, we get the right deformation up to terms vanishing due to the strong section condition. However, using the purely algebraic constraints derived in~\cite{Deser:2016qkw}, which are a weakening of the strong section condition, we even get exactly the same result as in~\eqref{eq:correctionsDFT}. The third relation of proposition~6.1 of~\cite{Deser:2016qkw} reads as
\begin{equation}
  \Bigl\lbrace \lbrace Q^2 X,Y\rbrace, Z\Bigr\rbrace_{[X,Y,Z]} =\, 2\theta^l\Bigl((\dpar^m X_l)(\dpar_m Y^k) Z_k\Bigr)_{[X,Y,Z]} = \,0\;,
\end{equation}
where the subscript $[X,Y,Z]$ denotes again the totally antisymmetrized sum over $X,Y,Z$. As this relation should hold for all degree~1 elements, choosing e.g.~an appropriate constant $Z_k$ gives the vanishing of the sum of the first and third $\alpha'$-correction in~\eqref{Cbracketdeformation}. The remaining terms are precisely the ones listed in~\eqref{eq:correctionsDFT}. To sum up, the deformation~\eqref{eq:def_Poisson_DFT} is the right one to interpret the  $\alpha'$-corrections encountered in Double Field Theory as a deformation of the Poisson structure used to construct the bilinear pairing and C-bracket. In~\cite{Deser:2014wva}, an attempt to get this deformed Poisson structure via deformation quantization of an inhomogeneous degree Poisson tensor was given. We leave it to future studies to investigate the properties of such deformations, e.g.~the failure of the Jacobi identity which might hint to non-associative structures.   

\section*{Acknowledgements}

The authors would like to thank Paolo Aschieri, Pedram Hekmati, Branislav Jurčo, Jim Stasheff, Satoshi Watamura and Martin Wolf for discussions. A.D. and C.S. are grateful to the Mathematical Institute of Charles University Prague for hospitality.

\appendices

\subsection{Relevant definitions}

In this section, we succinctly collect relevant definitions for the main text. For more detailed explanations, we refer to~\cite{Deser:2016qkw}. 

We always identify $\infty$-categorified Lie algebroids and Lie algebras (that is, Lie $\infty$-algebroids and Lie $\infty$-algebras), with $L_\infty$-algebroids or $L_\infty$-algebras. The latter are most efficiently defined in terms of N$Q$-manifolds.

In this paper, we define an {\em N-manifold} $\CM$ as an $\NN^*$-graded vector bundle $\CM\rightarrow M$ over some base manifold $M$. Locally, we thus have base coordinates of degree~0 on $M$ and fiber coordinates of degrees $1,2,...$ generating the $\NN$-graded algebra of functions $\CC^\infty(\CM)$. Endowing $\CM$ with a {\em homological vector field $Q$}, which is a vector field of degree~1 and satisfies $Q^2=0$, turns it into an {\em N$Q$-manifold} $(\CM,Q)$ and $(\CC^\infty(\CM),Q)$ becomes a differential graded algebra. 

By a {\em Lie $n$-algebroid}, we mean N$Q$-manifolds concentrated (i.e.\ non-trivial) in degrees $0,\dots,n$ and a {\em Lie $n$-algebra} is an N$Q$-manifold concentrated in degrees $1,\dots, n$. That is, a Lie $n$-algebra is a Lie $n$-algebroid, which is a vector bundle over a point.

An important example of an N-manifolds is the tangent bundle with fiber coordinates shifted in degree by 1, $T[1]M$. The algebra of functions on $T[1]M$ are the differential forms, $\CC^\infty(T[1]M)=\Omega^\bullet(M)$, and $T[1]M$ becomes an N$Q$-manifold with $Q$ the de Rham differential. 

Another example of an N-manifold is a grade-shifted Lie algebra $\frg[1]$. In coordinates $\xi^\alpha$ on $\frg[1]$, $\alpha=1,\dots,\dim(\frg)$ of degree~1, the most general vector field reads as $Q=-\tfrac12 f^\alpha_{\beta\gamma}\xi^\beta\xi^\gamma\der{\xi^\alpha}$ and $Q^2=0$ is equivalent to the Jacobi identity for the structure constants $f^\alpha_{\beta\gamma}$.

Finally, consider an N$Q$-manifold concentrated in degrees $1,\dots, n$ with coordinates $\xi_{(i)}^{\alpha_i}$ of degree~$i$. The vector field $Q$ will now contain a number of structure constants,
\begin{equation}
\begin{aligned}
 Q=&m^{\alpha_1}_{\alpha_2} \xi^{\alpha_2}_{(2)}\der{\xi^{\alpha_1}_{(1)}}+m^{\gamma_1}_{\alpha_1\beta_1}\xi^{\alpha_1}_{(1)}\xi^{\beta_1}_{(1)}\der{\xi^{\gamma_1}_{(1)}}+\\
 &+m^{\gamma_2}_{\alpha_1\beta_2}\xi^{\alpha_1}_{(1)}\xi^{\beta_2}_{(2)}\der{\xi^{\gamma_2}_{(2)}}+m^{\gamma_2}_{\alpha_1\beta_1\gamma_1}\xi^{\alpha_1}_{(1)}\xi^{\beta_1}_{(1)}\xi^{\gamma_1}_{(1)}\der{\xi^{\gamma_2}_{(2)}}+\dots~,
\end{aligned}
\end{equation}
which, in turn, define higher brackets on the homogeneously graded vector spaces $\CM_i$ of $\CM=\oplus_{i=1}^n \CM_i$, fully analogously to the above example of a Lie algebra. That is, introducing bases $\tau^{(i)}_{\alpha_i}$ on $\CM_i$, we have
\begin{equation}\label{eq:mu_from_Q}
\begin{aligned}
 \mu_1(\tau^{(i)}_{\alpha_i})&=m^{\alpha_{i-1}}_{\alpha_i}\tau^{(i-1)}_{\alpha_{i-1}}~,\\
 \mu_2(\tau^{(i)}_{\alpha_i},\tau^{(j)}_{\beta_j})&=m_{\alpha_i\beta_j}^{\gamma_{i+j-1}}\tau^{(i+j-1)}_{\gamma_{i+j-1}}~,
\end{aligned}
\end{equation}
etc. All these higher brackets are of degree~$-1$. It is customary to shift the degree of the $\CM_i$ by -1. This renders all $\mu_i$ totally graded antisymmetric\footnote{Note that for higher brackets $\mu_i$ with $i>3$, one has to introduce an ordering in the analogues of~\eqref{eq:mu_from_Q} since graded antisymmetry is not the same as graded symmetry after a shift in degree by one.} and shifts their degree to $i-2$. Also, the N$Q$-manifold $\frg[1]$ describing a Lie algebra is shifted to an ordinary vector space $\frg=\frg[0]$. In general, we obtain a graded vector space
\begin{equation}
 \sL=\oplus_{i=0}^{n-1} \sL_i\ewith \sL_i=\CM_{i+1}~.
\end{equation}
This is the traditional description of an $L_\infty$-algebra, cf.~\cite{Lada:1992wc}. The identity $Q^2=0$ reproduces the higher homotopy relations.

An N$Q$-manifold $\CM$ is a {\em symplectic N$Q$-manifold} $(\CM,Q,\omega)$ if it carries a symplectic form $\omega$ of homogeneous degree with respect to which $Q$ is an infinitesimal symplectomorphism: $\CL_Q\omega=0$. We call the $\NN$-degree of $\omega$ the {\em degree} of the symplectic N$Q$-manifold.

A simple example of a symplectic N$Q$-manifold is a metric Lie algebra $(\frg,\kappa)$. In the coordinates $\xi^\alpha$ introduced above, the symplectic form reads as $\omega=\tfrac12 \kappa_{\alpha\beta}\dd \xi^\alpha\wedge \dd \xi^\beta$, and $\CL_Q\omega=0$ is the compatibility condition of the metric $\kappa$ with the Lie bracket.

An important class of examples of symplectic Lie $n$-algebroids are the {\em Vinogradov algebroids} $\CV_n(M)=T^*[n]T[1]M$ over some manifold $M$, $n\in \NN^*$. Locally, they have coordinates $x^\mu,\xi^\mu,\zeta_\mu,p_\mu$ of degrees $0,1,n-1,n$ in which the canonical symplectic form reads as
\begin{equation}
 \omega=\dd x^\mu\wedge \dd p_\mu+\dd \xi^\mu\wedge \dd \zeta_\mu~.
\end{equation}
Clearly $\CV_n(M)$ is of degree~$n$. The homological vector field can be encoded in a Hamiltonian function as follows:
\begin{equation}
 Q=\{\CQ,-\}\ewith \CQ=\xi^\mu p_\mu~,
\end{equation}
where $\{-,-\}$ is the Poisson bracket of degree~$-n$ induced by $\omega$.

Less familiar than the above definitions may be the fact that any symplectic N$Q$-manifold (and therefore any symplectic $L_\infty$-algebroid) $(\CM,\omega)$ comes with an {\em associated Lie $n$-algebra} $\sL$ from antisymmetrized derived brackets~\cite{Roytenberg:1998vn,Fiorenza:0601312,Getzler:1010.5859}. For a symplectic N$Q$-manifold $(\CM,\omega)$ of degree~$n$, we have
\begin{subequations}\label{eq:ass_L_infty_structure}
\begin{equation}
\begin{aligned}
 \sL(\CM)&=~~\sL_0(\CM)&\leftarrow &~~\sL_1(\CM)&\leftarrow &\dots &\leftarrow &~~\sL_{n-1}(\CM)\\
 &=~~\CC^\infty_{n-1}(\CM)&\leftarrow &~~\CC^\infty_{n-2}(\CM) &\leftarrow &\dots &\leftarrow &~~\CC^\infty_0(\CM)~,
\end{aligned}
\end{equation}
where $\CC^\infty_i(\CM)$ are functions on $\CM$ of degree~$i$. The higher brackets $\mu_i$ are defined as follows. The lowest bracket is defined as a part of $Q$:
\begin{equation}
\mu_1(\ell)=\left\{\begin{array}{ll}
0 & \ell\in \CC^\infty_{n-1}(\CM)=\sL_0(\CM)~,\\
Q\ell & \mbox{else}~.\\
\end{array}\right.\\
\end{equation}
We call the remainder $q$,
\begin{equation}\label{def:delta}
 q(\ell)=\left\{\begin{array}{ll}
Q\ell & \ell\in \CC^\infty_{n-1}(\CM)=\sL_0(\CM)~,\\
0 & \mbox{else}~,\\
\end{array}\right.\\
\end{equation}
and use it to define totally antisymmetrized brackets
\begin{equation}\label{eq:L_infty_brackets}
\begin{aligned}
\mu_2(\ell_1,\ell_2)&=\tfrac12\big(\{q\ell_1,\ell_2\}\pm\{q\ell_2,\ell_1\}\big)~,\\
\mu_3(\ell_1,\ell_2,\ell_3)&=-\tfrac{1}{12}\big(\{\{q\ell_1,\ell_2\},\ell_3\}\pm \dots\big)~,
\end{aligned}
\end{equation}
\end{subequations}
where the sums run over all permutations and the signs are the obvious ones to conform with the symmetries of the $\mu_k$.

Note that there is also a graded Leibniz algebra structure of degree~$n+1$ on $\sL(\CM)$ given by the bracket
\begin{equation}\label{eq:Dorfman_bracket_gen}
 \lambda_2(\ell_1,\ell_2)=\{Q\ell_1,\ell_2\}~,
\end{equation}
which satisfies
\begin{equation}
 \lambda_2(\ell_1,\nu_2(\ell_2,\ell_3))=(-1)^{|\ell_1|+n+1}\lambda_2(\nu_2(\ell_1,\ell_2),\ell_3)+(-1)^{(|\ell_1|+n+1)(|\ell_2|+n+1)}\lambda_2(\ell_2,\nu_2(\ell_1,\ell_3))~.
\end{equation}
We often restrict the bracket $\lambda_2(\ell_1,\ell_2)$ to the {\em Dorfman bracket}
\begin{equation}\label{eq:def:Dorfman_bracket}
 \nu_2(\ell_1,\ell_2)=\{q\ell_1,\ell_2\}~,
\end{equation}
which is related to $\mu_2$ of the associated Lie $n$-algebra via
\begin{equation}
 \nu_2(\ell_1,\ell_2)=\mu_2(\ell_1,\ell_2)+\tfrac12 Q \{\ell_1,\ell_2\}
\end{equation}
for $\ell_1,\ell_2\in\sL_0$.

\bibliography{bigone}

\begin{thebibliography}{10}

\bibitem{Tseytlin:1990va}
A.~A.~Tseytlin,
{\em {Duality symmetric closed string theory and interacting chiral scalars},}
\href{http://dx.doi.org/10.1016/0550-3213(91)90266-Z}{Nucl. Phys. B {\bf 350}
  (1991) 395}.

\bibitem{Siegel:1993th}
W.~Siegel,
{\em Superspace duality in low-energy superstrings,}
\href{http://dx.doi.org/10.1103/PhysRevD.48.2826}{Phys. Rev. D {\bf 48} (1993)
  2826} [{\tt \href{http://www.arxiv.org/abs/hep-th/9305073}{hep-th/9305073}}].

\bibitem{Siegel:1993xq}
W.~Siegel,
{\em Two-vierbein formalism for string-inspired axionic gravity,}
\href{http://dx.doi.org/10.1103/PhysRevD.47.5453}{Phys. Rev. D {\bf 47} (1993)
  5453} [{\tt \href{http://www.arxiv.org/abs/hep-th/9302036}{hep-th/9302036}}].

\bibitem{Hull:2004in}
C.~M.~Hull,
{\em A geometry for non-geometric string backgrounds,}
\href{http://dx.doi.org/10.1088/1126-6708/2005/10/065}{JHEP {\bf 0510} (2005)
  065} [{\tt \href{http://www.arxiv.org/abs/hep-th/0406102}{hep-th/0406102}}].

\bibitem{Hull:2009sg}
C.~M.~Hull and R.~A.~Reid{--}Edwards,
{\em {Non-geometric backgrounds, doubled geometry and generalised T-duality},}
\href{http://dx.doi.org/10.1088/1126-6708/2009/09/014}{JHEP {\bf 0909} (2009)
  014} [{\tt \href{http://www.arxiv.org/abs/0902.4032}{0902.4032 [hep-th]}}].

\bibitem{Hull:2009mi}
C.~Hull and B.~Zwiebach,
{\em Double field theory,}
\href{http://dx.doi.org/10.1088/1126-6708/2009/09/099}{JHEP {\bf 0909}
  (2009)~99} [{\tt \href{http://www.arxiv.org/abs/0904.4664}{0904.4664
  [hep-th]}}].

\bibitem{Vaisman:2012ke}
I.~Vaisman,
{\em {On the geometry of double field theory},}
\href{http://dx.doi.org/10.1063/1.3694739}{J. Math. Phys. {\bf 53} (2012)
  033509} [{\tt \href{http://www.arxiv.org/abs/1203.0836}{1203.0836
  [math.DG]}}].

\bibitem{Vaisman:2012px}
I.~Vaisman,
{\em {Towards a double field theory on para-Hermitian manifolds},}
\href{http://dx.doi.org/10.1063/1.4848777}{J. Math. Phys. {\bf 54} (2013)
  123507} [{\tt \href{http://www.arxiv.org/abs/1209.0152}{1209.0152
  [math.DG]}}].

\bibitem{Deser:2014mxa}
A.~Deser and J.~Stasheff,
{\em {Even symplectic supermanifolds and double field theory},}
\href{http://dx.doi.org/10.1007/s00220-015-2443-4}{Commun. Math. Phys. {\bf
  339} (2015) 1003} [{\tt \href{http://www.arxiv.org/abs/1406.3601}{1406.3601
  [math-ph]}}].

\bibitem{Freidel:2017yuv}
L.~Freidel, F.~J.~Rudolph, and D.~Svoboda,
{\em {Generalised kinematics for Double Field Theory},}
{\tt \href{http://www.arxiv.org/abs/1706.07089}{1706.07089 [hep-th]}}.

\bibitem{Hitchin:2004ut}
N.~Hitchin,
{\em Generalized Calabi--Yau manifolds,}
\href{http://dx.doi.org/10.1093/qjmath/54.3.281}{Quart. J. Math. Oxford Ser.
  {\bf 54} (2003) 281} [{\tt
  \href{http://www.arxiv.org/abs/math.DG/0209099}{math.DG/0209099}}].

\bibitem{Hitchin:2005in}
N.~Hitchin,
{\em {Brackets, forms and invariant functionals},}
{\tt \href{http://www.arxiv.org/abs/math.DG/0508618}{math.DG/0508618}}.

\bibitem{Gualtieri:2003dx}
M.~Gualtieri,
{\em {Generalized complex geometry},} PhD thesis, Oxford (2003)
[{\tt \href{http://www.arxiv.org/abs/math.DG/0401221}{math.DG/0401221}}].

\bibitem{Roytenberg:0203110}
D.~Roytenberg,
{\em On the structure of graded symplectic supermanifolds and Courant
  algebroids,}
in: ``Quantization, Poisson Brackets and Beyond,'' ed.\ Theodore Voronov,
  Contemp. Math., Vol. 315, Amer. Math. Soc., Providence, RI, 2002
[{\tt \href{http://www.arxiv.org/abs/math.SG/0203110}{math.SG/0203110}}].

\bibitem{Deser:2016qkw}
A.~Deser and C.~Saemann,
{\em {Extended Riemannian geometry I: Local double field theory},}
{\tt \href{http://www.arxiv.org/abs/1611.02772}{1611.02772 [hep-th]}}.

\bibitem{Heller:2016abk}
M.~A.~Heller, N.~Ikeda, and S.~Watamura,
{\em {Unified picture of non-geometric fluxes and T-duality in double field
  theory via graded symplectic manifolds},}
\href{http://dx.doi.org/10.1007/JHEP02(2017)078}{JHEP {\bf 1702} (2017) 078}
  [{\tt \href{http://www.arxiv.org/abs/1611.08346}{1611.08346 [hep-th]}}].

\bibitem{Hohm:2017pnh}
O.~Hohm and B.~Zwiebach,
{\em {$L_{\infty}$ algebras and field theory},}
\href{http://dx.doi.org/10.1002/prop.201700014}{Fortsch. Phys. {\bf 65} (2017)
  1700014} [{\tt \href{http://www.arxiv.org/abs/1701.08824}{1701.08824
  [hep-th]}}].

\bibitem{Hohm:2011ex}
O.~Hohm and S.~K.~Kwak,
{\em Double field theory formulation of heterotic strings,}
\href{http://dx.doi.org/10.1007/JHEP06(2011)096}{JHEP {\bf 1106} (2011) 096}
  [{\tt \href{http://www.arxiv.org/abs/1103.2136}{1103.2136 [hep-th]}}].

\bibitem{Hohm:2013jaa}
O.~Hohm, W.~Siegel, and B.~Zwiebach,
{\em Doubled $\alpha'$-geometry,}
\href{http://dx.doi.org/10.1007/JHEP02(2014)065}{JHEP {\bf 1402} (2014) 065}
  [{\tt \href{http://www.arxiv.org/abs/1306.2970}{1306.2970 [hep-th]}}].

\bibitem{Hohm:2014xsa}
O.~Hohm and B.~Zwiebach,
{\em {Double field theory at order $\alpha'$},}
\href{http://dx.doi.org/10.1007/JHEP11(2014)075}{JHEP {\bf 1411} (2014) 075}
  [{\tt \href{http://www.arxiv.org/abs/1407.3803}{1407.3803 [hep-th]}}].

\bibitem{Hohm:2014eba}
O.~Hohm and B.~Zwiebach,
{\em {Green-Schwarz mechanism and $\alpha'$-deformed Courant brackets},}
\href{http://dx.doi.org/10.1007/JHEP01(2015)012}{JHEP {\bf 1501} (2015) 012}
  [{\tt \href{http://www.arxiv.org/abs/1407.0708}{1407.0708 [hep-th]}}].

\bibitem{Blumenhagen:2014iua}
R.~Blumenhagen and R.~Sun,
{\em {T-duality, non-geometry and Lie algebroids in heterotic double field
  theory},}
\href{http://dx.doi.org/10.1007/JHEP02(2015)097}{JHEP {\bf 1502} (2015) 097}
  [{\tt \href{http://www.arxiv.org/abs/1411.3167}{1411.3167 [hep-th]}}].

\bibitem{Garcia-Fernandez:2013gja}
M.~Garcia{--}Fernandez,
{\em {Torsion-free generalized connections and heterotic supergravity},}
\href{http://dx.doi.org/10.1007/s00220-014-2143-5}{Commun. Math. Phys. {\bf
  332} (2014)~89} [{\tt \href{http://www.arxiv.org/abs/1304.4294}{1304.4294
  [math.DG]}}].

\bibitem{Baraglia:2013wua}
D.~Baraglia and P.~Hekmati,
{\em Transitive Courant algebroids, string structures and T-duality,}
\href{http://dx.doi.org/10.4310/ATMP.2015.v19.n3.a3}{Adv. Theor. Math. Phys.
  {\bf 19} (2015) 613} [{\tt
  \href{http://www.arxiv.org/abs/1308.5159}{1308.5159 [math.DG]}}].

\bibitem{Sati:2009ic}
H.~Sati, U.~Schreiber, and J.~Stasheff,
{\em {Differential twisted String and Fivebrane structures},}
\href{http://dx.doi.org/10.1007/s00220-012-1510-3}{Commun. Math. Phys. {\bf
  315} (2012) 169} [{\tt \href{http://www.arxiv.org/abs/0910.4001}{0910.4001
  [math.AT]}}].

\bibitem{Fiorenza:2012tb}
D.~Fiorenza, H.~Sati, and U.~Schreiber,
{\em {Multiple M5-branes, string 2-connections, and 7d nonabelian Chern--Simons
  theory},}
\href{http://dx.doi.org/10.4310/ATMP.2014.v18.n2.a1}{Adv. Theor. Math. Phys.
  {\bf 18} (2014) 229} [{\tt
  \href{http://www.arxiv.org/abs/1201.5277}{1201.5277 [hep-th]}}].

\bibitem{Saemann:2017rjm}
C.~Saemann and L.~Schmidt,
{\em {The non-abelian self-dual string and the (2,0)-theory},}
{\tt \href{http://www.arxiv.org/abs/1705.02353}{1705.02353 [hep-th]}}.

\bibitem{Roytenberg:0112152}
D.~Roytenberg,
{\em Quasi-Lie bialgebroids and twisted Poisson manifolds,}
\href{http://dx.doi.org/10.1023/A:1020708131005}{Lett. Math. Phys. {\bf 61}
  (2002) 123} [{\tt
  \href{http://www.arxiv.org/abs/math.QA/0112152}{math.QA/0112152}}].

\bibitem{Deser:2014wva}
A.~Deser,
{\em {Star products on graded manifolds and $\alpha′$-corrections to Courant
  algebroids from string theory},}
\href{http://dx.doi.org/10.1063/1.4931137}{J. Math. Phys. {\bf 56} (2015)
  092302} [{\tt \href{http://www.arxiv.org/abs/1412.5966}{1412.5966
  [hep-th]}}].

\bibitem{Coimbra:2014qaa}
A.~Coimbra, R.~Minasian, H.~Triendl, and D.~Waldram,
{\em {Generalised geometry for string corrections},}
\href{http://dx.doi.org/10.1007/JHEP11(2014)160}{JHEP {\bf 1411} (2014) 160}
  [{\tt \href{http://www.arxiv.org/abs/1407.7542}{1407.7542 [hep-th]}}].

\bibitem{Bedoya:2014pma}
O.~A.~Bedoya, D.~Marques, and C.~Nunez,
{\em Heterotic $\alpha'$-corrections in double field theory,}
\href{http://dx.doi.org/10.1007/JHEP12(2014)074}{JHEP {\bf 1412} (2014) 074}
  [{\tt \href{http://www.arxiv.org/abs/1407.0365}{1407.0365 [hep-th]}}].

\bibitem{Gross:1985fr}
D.~J.~Gross, J.~A.~Harvey, E.~J.~Martinec, and R.~Rohm,
{\em {Heterotic string theory.~1. The free heterotic string},}
\href{http://dx.doi.org/10.1016/0550-3213(85)90394-3}{Nucl. Phys. B {\bf 256}
  (1985) 253}.

\bibitem{Gross:1985rr}
D.~J.~Gross, J.~A.~Harvey, E.~J.~Martinec, and R.~Rohm,
{\em {Heterotic string theory.~2. The interacting heterotic string},}
\href{http://dx.doi.org/10.1016/0550-3213(86)90146-X}{Nucl. Phys. B {\bf 267}
  (1986)~75}.

\bibitem{Bergshoeff:1981um}
E.~Bergshoeff, M.~de~Roo, B.~de~Wit, and P.~van~Nieuwenhuizen,
{\em {Ten-dimensional Maxwell--Einstein supergravity, its currents, and the
  issue of its auxiliary fields},}
\href{http://dx.doi.org/10.1016/0550-3213(82)90050-5}{Nucl. Phys. B {\bf 195}
  (1982)~97}.

\bibitem{Chapline:1982ww}
G.~F.~Chapline and N.~S.~Manton,
{\em {Unification of Yang--Mills theory and supergravity in ten dimensions},}
\href{http://dx.doi.org/10.1016/0370-2693(83)90633-0}{Phys. Lett. B {\bf 120}
  (1983) 105}.

\bibitem{Green:1984sg}
M.~B.~Green and J.~H.~Schwarz,
{\em {Anomaly cancellation in supersymmetric d=10 gauge theory and superstring
  theory},}
\href{http://dx.doi.org/10.1016/0370-2693(84)91565-X}{Phys. Lett. B {\bf 149}
  (1984) 117}.

\bibitem{Sati:2008eg}
H.~Sati, U.~Schreiber, and J.~Stasheff,
{\em $L_\infty$-algebra connections and applications to String- and
  Chern--Simons $n$-transport,}
in: ``Quantum Field Theory,'' eds. B. Fauser, J. Tolksdorf and E. Zeidler, p.
  303, Birkh{\"a}user 2009
[{\tt \href{http://www.arxiv.org/abs/0801.3480}{0801.3480 [math.DG]}}].

\bibitem{Gawedzki:1987ak}
K.~Gawedzki,
{\em {Topological actions in two-dimensional quantum field theories},}
Nonperturbative quantum field theory (Carg\`ese, 1987), 101--141, NATO Adv.
  Sci. Inst. Ser. B Phys., 185, Plenum, New York, 1988.

\bibitem{Freed:1999vc}
D.~S.~Freed and E.~Witten,
{\em Anomalies in string theory with D-branes,}
Asian J. Math {\bf 3} (1999) 819 [{\tt
  \href{http://www.arxiv.org/abs/hep-th/9907189}{hep-th/9907189}}].

\bibitem{Baez:2003aa}
J.~Baez and A.~S.~Crans,
{\em Higher-dimensional algebra VI: Lie 2-algebras,}
\href{http://tac.mta.ca/tac/volumes/12/15/12-15.pdf}{Th. App. Cat. {\bf 12}
  (2004) 492} [{\tt
  \href{http://www.arxiv.org/abs/math.QA/0307263}{math.QA/0307263}}].

\bibitem{Henriques:2006aa}
A.~Henriques,
{\em Integrating $L_\infty$-algebras,}
\href{http://dx.doi.org/10.1112/S0010437X07003405}{Comp. Math. {\bf 144} (2008)
  1017} [{\tt
  \href{http://www.arxiv.org/abs/math.CT/0603563}{math.CT/0603563}}].

\bibitem{Baez:2005sn}
J.~C.~Baez, D.~Stevenson, A.~S.~Crans, and U.~Schreiber,
{\em {From loop groups to 2-groups},}
\href{http://projecteuclid.org/euclid.hha/1201127333}{Homol. Homot. Appl. {\bf
  9} (2007) 101} [{\tt
  \href{http://www.arxiv.org/abs/math.QA/0504123}{math.QA/0504123}}].

\bibitem{Demessie:2016ieh}
G.~A.~Demessie and C.~Saemann,
{\em {Higher gauge theory with string 2-groups},}
\href{http://dx.doi.org/10.4310/ATMP.2017.v21.n8.a2}{Adv. Theor. Math. Phys.
  {\bf 21} (2017) 1895} [{\tt
  \href{http://www.arxiv.org/abs/1602.03441}{1602.03441 [math-ph]}}].

\bibitem{Severa:1998ac}
P.~Severa,
{\em Letter No.~3 to Alan Weinstein (1998),}
available at \href{http://sophia.dtp.fmph.uniba.sk/~severa/letters/}{\ttfamily
  http://sophia.dtp.fmph.uniba.sk/$\sim$severa/letters/}.

\bibitem{Garcia-Fernandez:2016ofz}
M.~Garcia{--}Fernandez,
{\em {Ricci flow, Killing spinors, and T-duality in generalized geometry},}
{\tt \href{http://www.arxiv.org/abs/1611.08926}{1611.08926 [math.DG]}}.

\bibitem{Jurco:2015bfs}
B.~Jurčo and J.~Vysoký,
{\em {Heterotic reduction of Courant algebroid connections and
  Einstein–-Hilbert actions},}
\href{http://dx.doi.org/10.1016/j.nuclphysb.2016.04.038}{Nucl. Phys. B {\bf
  909} (2016)~86} [{\tt \href{http://www.arxiv.org/abs/1512.08522}{1512.08522
  [hep-th]}}].

\bibitem{Jurco:2016emw}
B.~Jurco and J.~Vysoky,
{\em {Courant algebroid connections and string effective actions},}
in: ``Workshop on Strings, Membranes and Topological Field Theory,''
  Proceedings 2017, p.~211--265
[{\tt \href{http://www.arxiv.org/abs/1612.01540}{1612.01540 [math-ph]}}].

\bibitem{Jurco:2017gii}
B.~Jurco and J.~Vysoky,
{\em {Poisson-Lie T-duality of String Effective Actions: A New Approach to the
  Dilaton Puzzle},}
{\tt \href{http://www.arxiv.org/abs/1708.04079}{1708.04079 [hep-th]}}.

\bibitem{Hohm:2011si}
O.~Hohm and B.~Zwiebach,
{\em {On the Riemann tensor in double field theory},}
\href{http://dx.doi.org/10.1007/JHEP05(2012)126}{JHEP {\bf 1205} (2012) 126}
  [{\tt \href{http://www.arxiv.org/abs/1112.5296}{1112.5296 [hep-th]}}].

\bibitem{Geissbuhler:2013uka}
D.~Geissbuhler, D.~Marques, C.~Nunez, and V.~Penas,
{\em {Exploring double field theory},}
\href{http://dx.doi.org/10.1007/JHEP06(2013)101}{JHEP {\bf 06} (2013) 101}
  [{\tt \href{http://www.arxiv.org/abs/1304.1472}{1304.1472 [hep-th]}}].

\bibitem{Papadopoulos:2014mxa}
G.~Papadopoulos,
{\em {Seeking the balance: Patching double and exceptional field theories},}
\href{http://dx.doi.org/10.1007/JHEP10(2014)089}{JHEP {\bf 1410} (2014) 089}
  [{\tt \href{http://www.arxiv.org/abs/1402.2586}{1402.2586 [hep-th]}}].

\bibitem{Hohm:2013bwa}
O.~Hohm, D.~L{\"u}st, and B.~Zwiebach,
{\em {The spacetime of double field theory: Review, remarks, and outlook},}
\href{http://dx.doi.org/10.1002/prop.201300024}{Fortsch. Phys. {\bf 61} (2013)
  926} [{\tt \href{http://www.arxiv.org/abs/1309.2977}{1309.2977 [hep-th]}}].

\bibitem{Lada:1992wc}
T.~Lada and J.~Stasheff,
{\em {Introduction to sh Lie algebras for physicists},}
\href{http://dx.doi.org/10.1007/BF00671791}{Int. J. Theor. Phys. {\bf 32}
  (1993) 1087} [{\tt
  \href{http://www.arxiv.org/abs/hep-th/9209099}{hep-th/9209099}}].

\bibitem{Roytenberg:1998vn}
D.~Roytenberg and A.~Weinstein,
{\em {Courant algebroids and strongly homotopy Lie algebras},}
\href{http://dx.doi.org/10.1023/A:1007452512084}{Lett. Math. Phys. {\bf 46}
  (1998)~81} [{\tt
  \href{http://www.arxiv.org/abs/math.QA/9802118}{math.QA/9802118}}].

\bibitem{Fiorenza:0601312}
D.~Fiorenza and M.~Manetti,
{\em $L_\infty$ structures on mapping cones,}
\href{http://dx.doi.org/10.2140/ant.2007.1.301}{Alg. Numb. Th. {\bf 1} (2007)
  301} [{\tt
  \href{http://www.arxiv.org/abs/math.QA/0601312}{math.QA/0601312}}].

\bibitem{Getzler:1010.5859}
E.~Getzler,
{\em Higher derived brackets,}
{\tt \href{http://www.arxiv.org/abs/1010.5859}{1010.5859 [math-ph]}}.

\end{thebibliography}

\bibliographystyle{latexeu}

\end{document}